\documentclass[11pt,a4paper]{article}
\pdfoutput=1 %% activate this line when submit %%

\usepackage{jheppub}
\usepackage{amssymb,amsmath,amsthm,graphicx,mathrsfs,bbm}
\usepackage{latexsym,amscd,amsbsy,amsfonts,dsfont}
\usepackage{graphics, color}
\usepackage{subfigure}
\usepackage{bm}
\usepackage{epsfig}
\usepackage{textcomp}
\usepackage[utf8]{inputenc}
\usepackage{graphicx}
\usepackage{xcolor}
\usepackage{multirow}
\usepackage{makecell}

\allowdisplaybreaks[1]
\newcommand{\beq}{\begin{equation}}
    \newcommand{\eeq}{\end{equation}}
\newcommand{\beqa}{\begin{eqnarray}}
    \newcommand{\eeqa}{\end{eqnarray}}
\newcommand{\bal}{\begin{align}}
    \newcommand{\eal}{\end{align}}

\newcommand{\pd}{\partial}

\newcommand{\nn}{\nonumber}

\newcommand{\dalm}{\kern1pt\vbox{\hrule height 0.9pt\hbox{\vrule width 0.9pt\hskip 2.5pt\vbox{\vskip 5.5pt}\hskip 3pt\vrule width 0.3pt}\hrule height 0.3pt}\kern1pt}

\newcommand{\Or}[1]{\mathcal{O}\left(#1\right)}
\newcommand{\lp}{\left(}
\newcommand{\rp}{\right)}

\newcommand{\cR}{\mathcal{R}}

\newcommand{\hgfunc}[2]{ { \, {}_{#1}  F  {}_{#2} } }

\newcommand{\ord}[1]{{\cal O}\left( #1 \right)}
\newcommand{\nonum}{\nonumber\\}
\newcommand{\fr}[1]{\frac{1}{#1}}

\def\signK{K}

\definecolor{darkergreen}{rgb}{0.0, 0.6, 0.0}
\definecolor{shadeofblue}{rgb}{0,0.3,0.7}

\def\clock{{\count0=\time
        \divide\count0 60
        \ifnum\count0<10 0\fi\the\count0
        \multiply\count0 -60 \advance\count0 \time
        :\ifnum\count0<10 0\fi \the\count0
}}
\newcommand{\timestamp}{{\small\vbox{\hbox{\tt Draft by and for gladiators / \jobname.pdf}
            \hbox{\the\day/\the\month/\the\year, \clock
}}}}

\begin{document}
%---Title and Abstract ---------------------------------------------------------------
\title{The large $D$ effective theory of black strings in AdS}
\author[1,2]{David~Licht,}
\author[1,3,4]{Ryotaku~Suzuki,}
\author[1]{Benson~Way,}

\affiliation[1]{Departament de F\'{i}sica Qu\`{a}ntica i Astrof\'{i}sica, Institut de Ci\`{e}ncies del Cosmos,\\
Universitat de Barcelona, Mart\'{i} i Franqu\`{e}s, 1, E-08028 Barcelona, Spain}
\affiliation[2]{Department of Physics, Ben-Gurion University of the Negev, Beer-Sheva 84105, Israel.}
\affiliation[3]{Department of Physics, Osaka City University, Sugimoto 3-3-138, Osaka 558-8585, Japan.}
\affiliation[4]{Mathematical Physics Laboratory, Toyota Technological Institute,\\ Hisakata 2-12-1, Nagoya 468-8511, Japan.}
\emailAdd{david.licht@icc.ub.edu}
\emailAdd{sryotaku@toyota-ti.ac.jp}
\emailAdd{benson@icc.ub.edu}

\newcommand{\blue}{\color{blue}}
\newcommand{\red}{\color{red}}

\abstract{\noindent
We study black strings/funnels and other black hole configurations in AdS that correspond to different phases of the dual CFT in black hole backgrounds, employing different approaches at large $D$.  We assemble the phase diagram of uniform and non-uniform black strings/funnels and study their dynamical stability.  We also construct flowing horizons.  Many of our results are available analytically, though some are only known numerically.
}

%\pacs{}
\begin{flushright}
TTI-MATHPHYS-16
\end{flushright}

\maketitle

%---Main ---------------------------------------------------------------
%--------------------------------------------------------------
\section{Introduction}
%--------------------------------------------------------------

The black string is one of the most important exact solutions in higher dimensional general relativity.  It is simply the spacetime product of a Schwarzschild black hole and a circle.  But despite this almost trivial construction, the black string exemplifies much of the new gravtational physics that exists only in higer dimensions.  The black string and its associated solutions demonstrate that when the dimension $D\geq 5$, black holes can have non-spherical topology, are non-uniquely specified by their asymptotic charges~\cite{Gubser:2001ac,Wiseman:2002zc}, and exhibit instabilities that lead to a violation of cosmic censorship~\cite{Gregory:1993vy,Lehner:2010pn,Figueras:2022zkg}.

With the AdS/CFT correspondence providing one of the principal motivations for gravity in higher dimensions, it would be natural to investigate the phenomenology of black strings in AdS. 
One way to generate black strings in AdS is by foliating AdS spacetime with a lower-dimensonal AdS spacetime as follows:

%A big difference from the asymptotically flat case is that there are various configurations for the black strings corresponding to various foliations in AdS. For example, a simple implementation is the black string in the Ricci-flat slicing.
%\beq
%    ds^2 = \frac{L^2}{z^2}\left(dz^2 + ds_{\rm Schw}^2\right),
%\eeq
%where $ds_{\rm Schw}^2$ is the Ricci-flat metric for $D-1$-dimensional Schwarzschild black hole. 
%{\bf [some references more about this solution ? like Gregory(2000); on the singularity at $z\to \infty$]}
%However, we can consider another foliation such that each cross section also becomes AdS spacetime
\beq
   ds^2 = \frac{L^2}{\cos^2 z} \left(dz^2 + \frac{1}{L_{D-1}^2}ds_{AdS_{D-1}}^2\right)
\eeq
where $L$ and $L_{D-1}$ are the AdS radius for $AdS_D$ and $AdS_{D-1}$. The above metric is an AdS spacetime (i.e. a solution to the vacuum Einstein equation with a negative cosmological constant) if $ds_{AdS_{D-1}}^2$ is also an AdS spacetime.  By choosing $ds_{AdS_{D-1}}^2$ to be an AdS black hole, one can obtain a black string in AdS.

Because the string ends at two different positions on the AdS boundary ($z=\pm \pi/2$) it is natural to interpret the boundary CFT spacetime as one that contains a black hole. The AdS black string therefore provides an ideal setting for studying the physics of CFTs on black hole backgrounds.  From the CFT point of view, these black holes serve as heat baths that interact with the CFT.

As we shall see, AdS black strings can exhibit much of the same physics as black strings in flat space.  But being in AdS, this physics is now given a new interpretation within a CFT.  For instance, like flat black strings, AdS black strings can be non-unique and compete with other solutions with different topologies~\cite{Marolf:2019wkz}.  In the context of studying CFTs on black hole backgrounds via holography, the different bulk solutions are usually coined ``funnels,'' and ``droplets,'' depending on whether the bulk horizon is connected or disconnected, respectively~\cite{Hubeny:2009ru,Marolf:2013ioa}.  The competition between these different solutions within a thermodynamic ensemble is interpreted in the CFT as the competition between confining and deconfining phases.

Having two isolated black holes  on the AdS boundary, we can also encounter more unusual horizon configurations. By changing the relative temperature or rotation between the boundary black holes, one can generate bulk ``flowing'' horizons that are stationary, but non-Killing.  The CFT contains a steady-state flow with a nonzero entropy current.  Such solutions were first found in AdS~\cite{Fischetti:2012ps,Fischetti:2012vt,Figueras:2012rb}, though an analogous flat space example exists as well~\cite{Emparan:2013fha}.

Like flat black strings, AdS black strings can be unstable, and are also expected to violate cosmic censorship.  However, much remains unknown about the CFT interpretation of the naked singularity.  An early study of this system~\cite{Emparan:2021ewh} suggests that the effects of the singularity can potentially be large on boundary observables.

In this paper, we apply the large $D$ limit~\cite{Asnin:2007rw,Emparan:2013moa,Emparan:2020inr} towards the study of AdS strings.  In this limit, the near-horizon physics of black holes decouple from the asymptotic region, providing a set of effective equations that can be studied much more easily than the full Einstein equation at finite $D$~\cite{Emparan:2015hwa,Bhattacharyya:2015dva,Bhattacharyya:2015fdk}. So far, the large $D$ effective theory has been useful for studying various black holes/brane spacetimes. In particular, the dynamics of the asymptotically flat black string is understood at a nonlinear level~\cite{Emparan:2015hwa,Emparan:2015gva,Sorkin:2004qq,Suzuki:2015axa,Emparan:2018bmi}.   We will obtain a set of effective dynamical equations that govern the behaviour of AdS black strings. By focusing on the dynamics in the short wavelength of $\ord{1/\sqrt{D}}$ around $z=0$, these effective equations properly capture nonlinear dynamics of the AdS black strings~\cite{Emparan:2021ewh}\footnote{It is known that the nonlinear dynamics of rotating black holes is captured well by zooming in on the near-axis region, where the rotating solutions are described as Gaussian solutions, or {\it Gaussian blobs} in the effective theory of the black brane~\cite{Andrade:2018rcx,Andrade:2018yqu,Andrade:2019edf,Andrade:2018nsz,Licht:2020odx,Suzuki:2020kpx,Andrade:2020ilm}. Our approach in the AdS can be seen as a version of this {\it blob approximation}. }. In the present work, we more thoroughly investigate the various stationary solutions of these equations and study their linear stability.

\begin{figure}
    \centering
    \includegraphics[width=15cm]{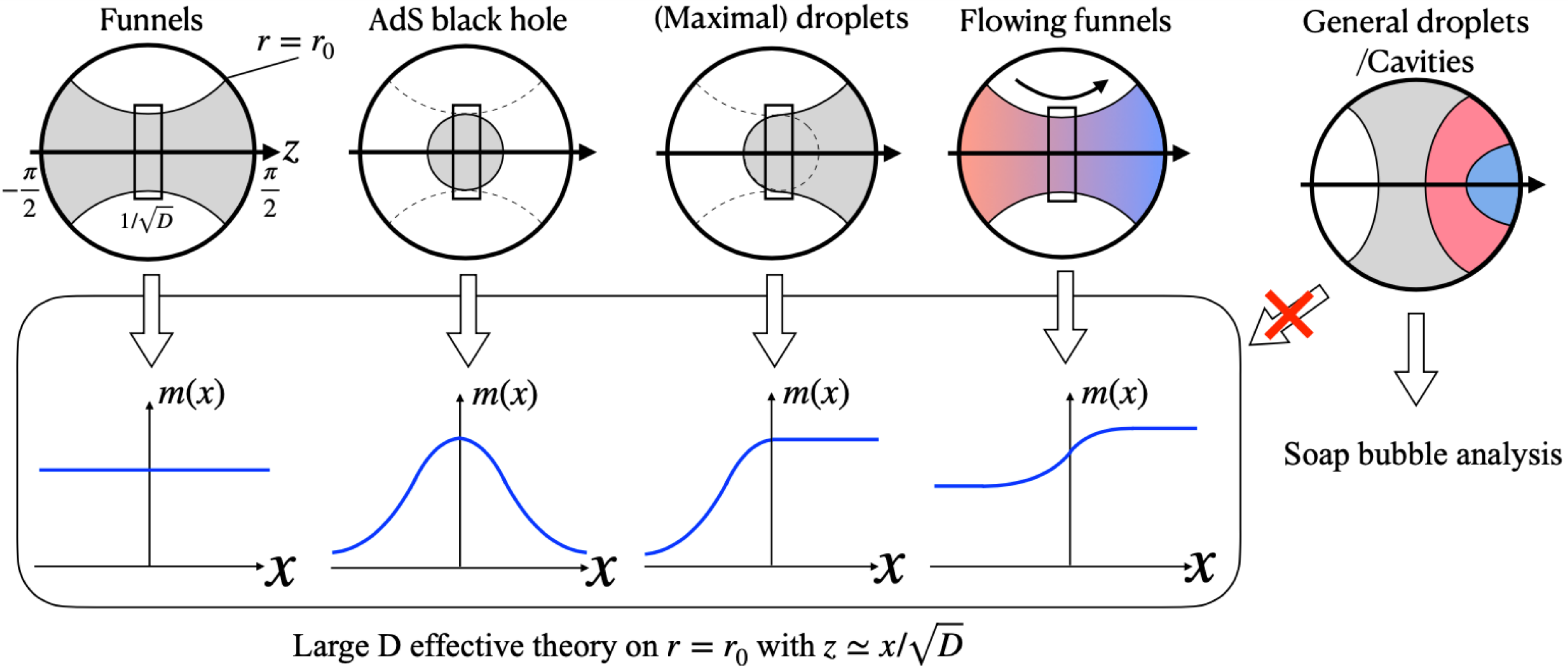}
    \caption{Sketch of the results from the large $D$ effective theory of AdS black strings for $K=1$. Black holes which are tangent to a $r$-constant surface at $z=0$ can be described as the solution with different mass density profiles in the effective theory around $z=0$. For other cases, we must find the embedding shape $r=r_0(z)$ by solving the soap bubble equation~\cite{Emparan:2015hwa}.}
    \label{fig:result-sketch}
\end{figure}
Using the large D effective theory approach, we obtain the following results (figure~\ref{fig:result-sketch}):
\begin{enumerate}
\item Possible static deformations of the black strings/funnels are studied. 
By solving the effective equation shown in ref.~\cite{Emparan:2021ewh}, we obtain not only deformed strings/funnels, but also an analytic Gaussian blob and droplets corresponding to different boundary conditions at AdS boundaries. We also study the thermodynamics and stability of the solutions. In particular, we analytically estimate the critical dimension (analogous to ref.~\cite{Sorkin:2004qq} in flat space) for weakly non-uniform strings/funnels, above which some non-uniform solutions are thermodynamically preferred.
\item We also study general horizon shapes embedded in global AdS that cannot be described by the large $D$ effective theory on the uniform strings/funnels.
The embedded horizon should satisfy the embedding condition, which is called the soap bubble equation. We obtain droplets, funnels and AdS black holes by numerically solving the soap bubble equation~\cite{Emparan:2015hwa}.
\item We show the large $D$ effective theory on the uniform strings/funnels also admits non-static but steady-state configurations, or flowing solutions, by imposing slightly different temperatures at AdS boundaries.
\end{enumerate}

%All of the above results stem from a set of effective equations that can be described as small (in $1/D$) flucutations about a known uniform string solution.  This uniform string at large $D$ solves a ``soap-bubble" equation for the embedding of its horizon in AdS.  The fact that this soap-bubble equation is solved analytically by uniform strings enables us to easily obtain the dynamical effective equations that govern the small horizon fluctuations.

%Naturally, there are other embedding solutions to the soap-bubble equation.  In this work, we also obtain some of these embeddings that correspond to droplet solutions, many of which lie outside of the set of solutions described by the dynamical effective theory about uniform strings.  But these droplet embeddings are generally only known numerically, so we cannot obtain their corresponding effective dynamical equations as easily.

In the following section, we review the large $D$ effective equations for AdS strings first obtained in \cite{Emparan:2021ewh}.  Then, in section~\ref{sec:thermo}, we explain the thermodynamics of black holes within this effective theory.  We then study the thermodynamic and stability of some exact solutions in section~\ref{sec:thermostabexact}, followed by some numerical solutions in section~\ref{sec:numericalphase}.  In section~\ref{sec:soapbubble}, we present and solve the soap-bubble equations for AdS horizon embeddings and connect some of these solutions to the exact solutions of the effective theory. In section~\ref{sec:flowingfunnels}, we construct solutions with flowing horizons at large $D$.  We summarize our findings and provide some concluding remarks in section~\ref{sec:summary}.

%--------------------------------------------------------------
\section{Large \texorpdfstring{$D$}{D} effective theory near the neck}\label{sec:largeDeffectivetheory}
%--------------------------------------------------------------

Here, we obtain the large $D$ effective equations that describe the dynamics of AdS black strings.  As is typical of the large $D$ limit, the Einstein equation reduces to a set of membrane equations that govern small fluctuations of the horizon.  These equations decouple the horizon from its environment, and hence only depend on time, and the spatial directions along the horizon.

However, it turns out that the the long and short wavelengths along the string (relative to $D$) are also be separated.  We therefore have two sets of effective dynamical equations: one that governs short wavelengths of $O(1/\sqrt{D})$, and another that governs long $O(1)$ wavelengths.  Since the short wavelength sector contains instabilities, it contains more interesting dynamics, and is the focus of this section. The long wavelength effective equations can be found in Appendix~\ref{app:longwavelength}.

\subsection{Foliations in AdS}
To form a black string in AdS, one can foliate AdS with AdS slices\footnote{One can similarly foliate AdS with flat, or deSitter slices, but the resulting black strings are singular.}.  That is, the following metric solves the vacuum Einstein equation with negative cosmological constant:
\begin{equation}\label{eq:foliation}
    ds^2=\frac{L^2}{\cos^2 z}\left(dz^2+\frac{1}{L_{D-1}^2}ds^2_{AdS_{D-1}}\right)\,,
\end{equation}
where $L$ is the AdS length scale and $ds^2_{AdS_{D-1}}$ is also an AdS metric. To obtain a black string, we can choose the Schwarzschild-AdS metric
\begin{equation}\label{eq:string}
    ds^2_{AdS_{D-1}}=L_{D-1}^2\left(-f(r)d\tau^2+\frac{dr^2}{f(r)}+r^2d\Sigma^2_{D-3}\right)\,,\qquad f(r)=\signK +r^2-\frac{\mu}{r^{D-4}}\;,
\end{equation}
where $d\Sigma^2$ is the metric for a maximally symmetric space with constant curvature $\signK \in\{1,0,-1\}$.  Written this way, the full metric is now independent of $L_{D-1}$.  The case $\signK =1$ best describes a black string, though we will keep $\signK $ general.  For $\signK =1$ and $0<\mu\ll 1$, the region with small $z$ and $r$ closely resembles a black string in flat space\textbf{}.

The boundary metric is obscured by the fact that both $z\to\pm\pi/2$ and $r\to\infty$ lie at the AdS boundary.  Therefore, consider the coordinate transformation
\begin{equation}
    r\to\frac{\rho\,C_\signK (\xi)}{\sqrt{1+\rho^2S_\signK (\xi)^2}}\;,\qquad \tan(z)\to\rho \,S_\signK (\xi)\;,
\end{equation}
where
\begin{subequations}\label{trigfcns}
\begin{align}
    &S_{1}(\xi)=\sin\xi\;,\qquad S_{0}(\xi)=\xi\;,\qquad S_{-1}(\xi)=\sinh\xi\;,\\
    &C_{1}(\xi)=\cos\xi\;,\qquad C_{0}(\xi)=1\;,\qquad C_{-1}(\xi)=\cosh\xi\;,\\
    &T_{1}(\xi)=\tan\xi\;,\qquad T_{0}(\xi)=\xi\;,\qquad T_{-1}(\xi)=\tanh\xi\;.
\end{align}
\end{subequations}
With these new coordinates and $\mu=0$, the metric \eqref{eq:foliation} takes the familiar form
\begin{equation}
    ds^2=L^2\left[-(\signK +\rho^2)d\tau ^2+\frac{d\rho^2}{\signK +\rho^2}+\rho^2\left(d\xi^2+C_\signK (\xi)^2d\Sigma^2_{D-3}\right)\right]\;,\label{eq:bg-vacuum-AdS}
\end{equation}
which is just the metric for vacuum AdS.  Now applying the transformation when $\mu\neq0$, we can then take the $\rho\to\infty$ limit to extract the boundary metric, which can be written in the form
\begin{equation}\label{eq:bdrymetric}
    ds^2_\partial=-\left[1-\mu\, S_\signK (\xi)^2|T_\signK (\xi)|^{D-4}\right]d\tau^2+\frac{d\xi^2}{1-\mu\,S_\signK (\xi)^2|T_\signK (\xi)|^{D-4}}+C_\signK(\xi)^2d\Sigma^2_{D-3}\;.
\end{equation}
This boundary metric contains two horizons, is $\mathbb Z_2$ symmetric about $\xi=0$, and is smooth for even $D$.

\subsection{Large \texorpdfstring{$D$}{D} limit of the AdS string}
Let us now write the AdS black string in a form that is more amenable to taking a large $D$ expansion.  We begin by taking the AdS black string and replacing $\mu$ with the horizon radius $r_0$, and moving to Eddington-Finkelstein coordinates
\begin{equation}\label{eq:AdS-blackstring-sol}
    ds^2=\frac{L^2}{\cos^2 z}\left(dz^2-f(r)d\tau^2+2dtdr+r^2d\Sigma^2_{D-3}\right)\,,\qquad f(r)=\signK +r^2-\frac{r_0^{D-4}(\signK +r_0^2)}{r^{D-4}}\;.
\end{equation}
For convenience, we set
\begin{equation}
    n = D-4\;,
\end{equation}
and will expand in $1/n$.  We define the large $D$ radial coordinate, and a short wavelength variable
\begin{equation}\label{eq:coordtransf}
    \frac{r}{r_0}\to e^{\frac{\rho}{n}}\;,\qquad z \to \frac{x}{\sqrt{n}}, %\cos z\to 1-\frac{x^2}{2n}\;,
\end{equation}
which motivates the ansatz
\begin{equation}\label{eq:metans}
    ds^2=\frac{L^2}{\cos^2z}\left[-\left(1+\frac{\signK }{r_0^2}\right)Adt^2+\frac{2e^{\frac{\rho}{n}}}{n}U dtd\rho+\frac{G^2}{n }\left(dx-\frac{C}{G^2}dt\right)^2+r_0^2e^{\frac{2\rho}{n}}d\Sigma_{n+1}\right]\;,
\end{equation}
where $A$,$C$,$U$,$G$ are functions of $(t,x,\rho)$, and $\cos z$ depends on $x$ according to \eqref{eq:coordtransf}.  With this ansatz, the black string is recovered with
\begin{equation}\label{eq:ansatzonstring}
    A=\frac{e^{\frac{2\rho}{n}}+\signK /r_0^2}{1+\signK /r_0^2}-e^{-\rho}\;,\qquad C=0\;,\qquad U=G=1\;.
\end{equation}
Note that in this ansatz we have rescaled the time coordinate by $t\to r_0^{-1} t$ for convenience.

\paragraph{Effective equations.}
From here, we can proceed with expanding the Einstein equation and solving for the metric functions order by order in $1/n$.  We impose boundary conditions also order by order in $1/n$.  For the first two orders in the expansion, we require that at $\rho\to\infty$, we get
\begin{equation}
    A= 1+\frac{2\rho}{\left(1+\signK /r_0^2\right)n}+O(n^{-2})\,,\quad G=1+O(n^{-2})\,,\quad C=O(n^{-2})\,,\quad U=1+O(n^{-2})\;.
\end{equation}
These conditions recover the asymptotic AdS boundary, and are consistent with the black string solution given by \eqref{eq:ansatzonstring}.

The leading order solution of the Einstein equation yields
\begin{align}\label{eq:ACHu}
A&=1-e^{-\rho}m(t,x)\,,& C&=e^{-\rho}p(t,x)\,,\\
G&=1+\frac{e^{-\rho}p^2(t,x)}{2(1+\signK /r_0^2) m(t,x)}\frac{1}{n}\,,& U&=1-\frac{e^{-\rho}p^2(t,x)}{2(1+\signK /r_0^2) m(t,x)}\frac{1}{n}\;.
\end{align}
The integration functions $m$ and $p$ must satisfy the effective equations
\begin{subequations}\label{eq:effeqns}
\begin{align}
    \partial_t m + (\partial_x +x) (p-\partial_x m)&=0\,,\\
    \partial_t p-(\partial_x +x)\left(\partial_x p-\frac{p^2}{m}\right) + p-\left(1+\frac{\signK }{r_0^2}\right)\partial_x m&=0\,.
\end{align}
\end{subequations}
The black string is recovered for the solution $m=m_0$, $p=0$.

At this point, we see that the equations depend only on the combination $\signK /r_0^2$, and not on $\signK $ and $r_0$ individually.  We therefore define
\begin{equation}
    \alpha\equiv\frac{\signK }{r_0^2}\;,\label{eq:def-alpha}
\end{equation}

From \eqref{eq:coordtransf}, we can trace the origin of derivative terms to the spatial divergence and Laplacian:
\beq\label{eq:divx}
\sqrt{n}\cos^n z\pd_z \cos^{-n}z \simeq e^{-x^2/2}\pd_x\, e^{x^2/2} = \pd_x +x
\eeq
\beq\label{eq:lapx}
n\cos^n z\pd_z\lp \cos^{-n}z\pd_z\rp\simeq e^{-x^2/2}\lp \pd_x\, e^{x^2/2}\pd_x\rp=\pd_x^2+x\pd_x
\eeq

\paragraph{Hydrodynamic form of the equations.} Define a velocity $v(t,x)$ such that
\beq
p=mv +\partial_x m\,.
\eeq
The equations take the form
\beq\label{eq:dtm}
\pd_t m + \lp \pd_x +x\rp \lp m v\rp=0\,,
\eeq
\beq\label{eq:dtmv}
\partial_t (mv)-\alpha\pd_x m+\lp \pd_x +x\rp \lp m v^2 -2 m\pd_x v-m\pd_x^2\ln m\rp=0\,.
\eeq
The first equation can be understood as the conservation of the mass flux if we rewrite it as
\beq
 \pd_t \lp e^{x^2/2} m\rp + \pd_x \lp e^{x^2/2} mv \rp =0.
\eeq
Later, we will see this actually equivalent to the conservation of the Brown-York energy flux with the spacial volume factor $\propto \cos^{n+2}z \sim e^{x^2/2}$. In contrast, eq.~\eqref{eq:dtmv} cannot be written in a conservative form in curved spacetime, since the pressure term ($-\alpha\partial_x m$) uses the usual gradient and not by the gradient with the warp factor like other terms.  This is natural since we do not have translation symmetry along the string.

For $\signK =1$ (that is, $\alpha=1/r_0^2$), one can recover the equation for black strings in flat space (not to be confused with the AdS black brane with one active direction) by focusing on the small central region $r_0|x|\ll 1$. Let us make following rescaling
%\beq
%t = r_0^{2}\bar{t} ,\quad x = r_0 \bar{x},\quad v = r_0^{-1} \bar{v},\quad p=r_0^{-1} \bar{p}. \label{eq:coordinate-AF}
%\eeq
\beq
t = \alpha \,\bar{t} ,\quad x = \alpha^{-1/2}\, \bar{x},\quad v = \alpha^{1/2}\, \bar{v},\quad p= \alpha^{1/2}\, \bar{p}. \label{eq:coordinate-AF}
\eeq
Then, in the limit $\alpha\to \infty \, (r_0\to 0)$, we recover the equations for the asymptotically (Kaluza-Klein) flat black string. %Note that the divergence of the pressure term is not modified by the cosmological constant, and this precludes the possibility of writing this equation as a conservation equation (in a curved geometry).

We also obtained the $1/n$-corrections to metric solutions and effective equations up to $\ord{1/n^2}$.
Since they are lengthy to show in the paper, we instead only present the some important results with them.

\subsection{Symmetries}
\paragraph{Scale invariance}
Eqs.~\eqref{eq:dtm} and \eqref{eq:dtmv} are invariant under the rescaling
\beq\label{eq:scaling}
m \to \lambda m,\quad p \to \lambda p.
\eeq
This can be traced back to the scaling degree of freedom in $r_0$.
%Actually, this scaling law is present even when including higher-order corrections in $1/n$. can be enhanced to include higher order corrections by defining $(m,p)$ up to higher order as \BW{I don't understand what this means.}
%\beq
% A(\sR=m)=0,\quad C(\sR=m) = \frac{p}{m}.
%\eeq
%Then, the corrected effective equation has the following scaling invariance \BW{Shall we exchange $\ell\to1/r_0$?}
%\beq
% \ell \to \lambda^{\frac{1}{n}}\ell,\quad
% t \to \lambda^{-\frac{1}{n}}t,\quad
% m \to \lambda m,\quad p \to \lambda p.
%\eeq
%Note that the $x$ coordinate is dimensionless and hence is not involve in the scaling.

\paragraph{Boost-like symmetry}
The effective theory inherits a boost symmetry which comes from the embedding of $AdS_D$ into $\textrm{Mink}_{2,D-1}$. The detailed derivation is presented in appendix \ref{app:embeddingAndBoostingOfAdS}.
Due to the boost symmetry, the following $m(t,x)$ and $v(t,x)$ solve the effective equation (\ref{eq:dtm}) and (\ref{eq:dtmv}), provided that $\bar{m}(t,x)$ and $\bar{v}(t,x)$ are also solutions:
\begin{align}
    &m(t,x) = \bar{m}(t,x+V(t))e^{\frac{V(t)^2}{2}+V(t) x},\quad
    v(t,x) = \bar{v}(t,x+V(t))- V'(t)\label{eq:boostsym}
\end{align}
where the boost function satisfies
\begin{align}
    V''(t) + \alpha V(t) = 0\;,
\end{align}
which is solved by
\begin{equation}
    V(t) = A\,S_{\signK}\left(\tfrac{t}{r_0}+B\right)
    ,\quad A,B\ :\ {\rm constant}\;,\label{eq:boostsym-Vt}
\end{equation}
where $S_K$ is given by \eqref{trigfcns}.  Note that eq.~(\ref{eq:boostsym}) looks simpler when it is rewritten in terms of the mass density $m e^{x^2/2}$
\begin{align}
    m(t,x) e^\frac{x^2}{2} = \bar{m}(t,x+V(t)) e^{\frac{1}{2}(x+V(t))^2}.
\end{align}
 For $\alpha=0$, this gives the same boost symmetry in the asymptotically flat case~\cite{Andrade:2018nsz}.
%
%%%%%%%%%%

%%%%%%%%%%
%
%\subsection{Physical quantities: Mass, entropy and free energy}

\subsection{Static configurations}
On solutions that are static, the equations of motion take a simpler form.  First set $\pd_t m=0$ and $v=0$ (so then $p=\pd_x m$) so that the equations reduce to\footnote{Note that the static condition comes from $T_{tz}=0$. With $1/n$-correction, this leads to $p=\partial_x m + \ord{1/n}$.}
\beq\label{eq:staticeq-m-3rd}
\alpha\, \pd_x m+\lp \pd_x +x\rp \lp m\,\pd_x^2\ln m\rp=0\,.
\eeq
Now introduce
\begin{equation}
\cR(x)=\ln m(x)
\end{equation}
so the equation becomes
\beq
\alpha\cR'+\cR'\cR''+\cR'''+x\cR''=0
\eeq
which can be easily integrated once to yield the second-order ODE
\beq\label{eq:staticeq}
\cR''+\frac12\cR'^2+\lp \alpha-1\rp \lp\cR-\cR_0\rp +x\cR'=0\,.
\eeq
The integration constant $\cR_0$ amounts to a constant rescaling of $m$.

\paragraph{Asymptotic behavior}
Here we estimate the asymptotic behavior of the static solution at $x\to \pm \infty$, when the solution approaches a fixed boundary value $\cR(x) \simeq \cR_0 + \delta \cR(x)$. By solving the linearized equation, we obtain the general solution for non-integer $\alpha$ in terms of Hermite polynomials
\beq
\delta \cR(x) = A\,  e^{-\frac{x^2}{2}} H_{\alpha-2}\left(\frac{|x|}{\sqrt{2}}\right)
%+ B \,  e^{-\frac{x^2}{2}}  {}_1F_1\left(1-\frac{\alpha}{2};\frac{1}{2};\frac{x^2}{2}\right).
+ B \,   e^{-\frac{x^2}{2}} H_{\alpha-2}\left(-\frac{|x|}{\sqrt{2}}\right)
\eeq
Thus, for $x\to \infty$, we obtain
\beq
%\delta \cR(x) \simeq A \left(\sqrt{2}|x|\right)^{\ell^2-2}e^{-\frac{x^2}{2}} +   B \frac{ \sqrt{\pi}}{\Gamma\left(1-\frac{\ell^2}{2}\right)}\left(\frac{|x|}{\sqrt{2}}\right)^{-\ell^2+1}.\label{eq:staticeq-bdry-bhv}
\delta \cR(x) \simeq A \left(\sqrt{2}|x|\right)^{\alpha-2}e^{-\frac{x^2}{2}} +   B \frac{ \sqrt{\pi}}{\Gamma\left(2-\alpha\right)}\left(\frac{|x|}{\sqrt{2}}\right)^{1-\alpha}.\label{eq:staticeq-bdry-bhv}
\eeq
If $B=0$, the deformation exponentially decays to the uniform value. This is actually consistent with the perturbative analysis in the $z$-coordinate~\cite{Marolf:2019wkz} (see Appendix~\ref{app:marolfsantos}).

%The another choice $B\neq0$, however, is also physical, which is connected to the another effective theory with long scaling. This will be elaborated in later section.\RS{We need more study in this aspect}

%--------------------------------------------------------------
\section{Thermodynamics}\label{sec:thermo}
%--------------------------------------------------------------

Here we study the thermodynamic properties of the effective theory.  Recall that the boundary metric can be described as having two black holes.  These black holes can exchange mass and entropy with the CFT, so these quantities are no longer conserved.  In the bulk, these conservation laws are broken due to the fact that the horizon is non-compact.

Nevertheless, there is still a second law of thermodynamics, which is governed by the free energy, which always monotonically decreases.

\subsection{Mass and entropy current, local temperature}
\paragraph{Mass current}
The Brown-York quasi-local tensor can be obtained in the same way as in black strings in flat space. Since we do not have spacial translation symmetry, the only conserved quantity is the mass density current which is given in terms of the Brown-York tensor,
\begin{equation}
    \sqrt{-g_{n+3}} T^\mu{}_t \partial_\mu = \frac{\sqrt{n}\Omega_{n+1}}{2} \frac{1+\alpha}{r_0}\left(r_0 L \right)^{n+2}
    \left(\rho_M \partial_t + j_M \partial_x\right)
\end{equation}
which is defined on the constant $\rho$ surface\footnote{This is not the conformal boundary geometry at $r=\infty$, but rather the matching region within the near horizon region where $0 \ll {\rho} \ll n$.}
\begin{equation}
    ds^2_{n+3} = \frac{L^2}{\cos^2(x/\sqrt{n})}\left(\frac{dx^2}{n}-(\alpha+1)dt^2+r_0^{2}{e}^{2\rho/n}d\Omega_{n+1}^2\right)+\ord{\rho,e^{-\rho}}.
\end{equation}
Up to $\ord{1/n}$, the mass density $\rho_M$ and flux $j_M$ are given by
\begin{equation}
    \rho_M =e^{\frac{x^2}{2}} m+\frac{e^{\frac{x^2}{2}}}{n} \left(m \left(\frac{2 \log
        m}{\alpha+1}+\frac{x^4}{12}+x^2+1\right)-\frac{\left(\partial_x p+x
        p\right) (\log m+2)}{\alpha+1}\right)
\end{equation}
and
\begin{align}
    j_M=e^{\frac{x^2}{2}} \left(p -\partial_x m \right)
    +&\frac{e^{\frac{x^2}{2}}}{n(\alpha+1)} \left[\frac{1}{12} \left(\left(\alpha+1\right) x^4+12 \left(\alpha+1\right) x^2+24\right)
    \left(p -\partial_x m \right)\right.+\nonum
    &   \left.\qquad -\frac{\left(\partial_x p +x p \right)
        \left(p -\partial_x m \right)}{m}+\frac{p^2\partial_x m }{2 m ^2}\right.+\nonum
    &\left.\qquad+\left(\frac{p^2\partial_x m }{m ^2}
    -2 \partial_x m -\frac{\left(2 \partial_x p +x p \right)
        p }{m}+x \partial_x p +\partial_x^2p +p \right) \log
    m\right].
\end{align}
Using the effective equation to next-to-leading order, the following conservation law holds up to the relevant order,
\begin{equation}
    \partial_t \rho_M + \partial_x j_M = 0.
\end{equation}
However, the asymptotic behavior of the static solution~\eqref{eq:staticeq-bdry-bhv} indicates that, at leading order,
\beq
\rho_M = e^\frac{x^2}{2} m \sim |x|^{\alpha-2}, \quad (|x|\gg 1)
\eeq
which does not provide convergent integration for $\alpha \geq 1$.  Similarly, the flux $j_M$ does not generally vanish as $x\to \pm\infty$.  That is, the total mass is formally infinite, and its conservation via the mass flux is not guaranteed.

Note that the regularization of the quasi-local tensor has an ambiguity of adding a functional degree of freedom $\phi(t,x)$,
\beq
\rho_M \to \rho_M + \partial_x \phi,\quad j_M \to j_M - \partial_t \phi.
\eeq
In the asymptotically flat case, this freedom in $\phi$ never changes the total mass. However, in the current setup, different $\phi$ will change the boundary behavior at $x\to \pm \infty$. Later, we will see how this ambiguity has consequences for the definition of free energy.

\paragraph{Entropy current and local temperature}
The entropy and local temperature are defined on the local event horizon~\cite{Bhattacharyya:2008xc}.
The entropy current is given by
\begin{equation}
    J_s^\mu \partial_\mu = \frac{r_0 \Omega_{n+1}}{\sqrt{n}} \left(r_0 L\right)^{n+2} \left(\rho_S\partial_t+j_S \partial_x \right)
\end{equation}
where
\begin{align}
    \rho_S =e^{\frac{x^2}{2}} m +  \frac{e^{\frac{x^2}{2}}}{n}&\left[\frac{2 \left(x \partial_x m +\partial_x^2 m -\partial_x p -x
        p \right)}{\alpha+1}-\frac{p^2-4 p\partial_x m  +2 (\partial_x m)^2}{2 \left(\alpha+1\right) m }\right.\nonum
    &\qquad\qquad\qquad\qquad\qquad\qquad\qquad\qquad\qquad\left.+\left(\frac{x^4}{12}+x^2+\log m\right) m \right],
\end{align}
and
\begin{align}
    & j_S=e^{\frac{x^2}{2}}\left(p -\partial_x m \right)
    +\frac{e^{\frac{x^2}{2}}}{n(\alpha+1)} \left[\frac{\left(\left(\alpha+1\right) x^4+12 \left(\alpha+1\right) x^2+24\right)
        \left(p - \partial_x m \right)}{12}-2  \partial_x^3 m +2  \partial_x^2 p\right.\nonum
    &\quad \left.+\left(\alpha+1\right) \left(p -2
    \partial_x m \right) \log m +\frac{2 \left(p - \partial_x m +x
        m  \right) \left( \partial_x p - \partial_x^2 m \right)}{m }
    -\frac{p^3 -4    (\partial_x m) ^2 p +2  (\partial_x m) ^3}{2 m ^2} \right].
\end{align}
Using the effective equation, we can deduce the second law
\begin{equation}
    \partial_t  \rho_S + \partial_x j_S = \frac{2}{n(\alpha+1)}m(\partial_x v )^2 \geq 0,
\end{equation}
where $v:=(p-\partial_x m)/m$.
For the same reason as the mass current, the total entropy is not a well-defined quantity.

The local temperature is given by the surface gravity of the local event horizon as
\begin{equation}
    { \boldsymbol \kappa} = \frac{n}{2r_0}\left(1+\alpha\right)\kappa
\end{equation}
where $\kappa$ is given up to next to leading order
\begin{align}\label{eq:kappa-nlo}
    \kappa =1 + \frac{1}{(\alpha+1)n}&\left(2-\left(\alpha-1\right) \log m +\frac{\partial_x p +x p-2 \partial_x^2 m  }{m }+\right.\nonum
    &\left.\qquad\qquad+\frac{3 p ^2-4 p\partial_x m  +2 \partial_x m  \left(\partial_x m -2 x
        m \right)}{2 m ^2}\right).
        \end{align}
In the static case, $\partial_x \kappa =0$ gives the soap bubble equation equivalent to the effective equation.
The boundary condition $m = m_0+\ord{e^{-x^2/2}},\ p = \ord{e^{-x^2/2}}$ for $x \to \pm\infty$
fixes the boundary temperature
\begin{equation}
    \kappa \to \kappa_0 =  1 + \frac{2-(\alpha-1)\log m_0}{n (\alpha+1)}.\label{eq:bdry-temp}
\end{equation}
This reproduces the temperature of the uniform AdS string~\eqref{eq:adsstring} with $r_H=r_0 m_0^{1/n}$,
\begin{equation}
    r_0{\boldsymbol \kappa} = \frac{n}{2}\left(r_H+\fr{r_H}\right)+r_H,
\end{equation}
where one should note that we have rescaled the time coordinate by $r_0^{-1}$.

\subsection{Free energy}
Neither the total mass nor total entropy are well-defined physical quantities in the effective theory due to the non-vanishing boundary fluxes and non-convergence of the integration.  Nevertheless, one can see that the free energy defined under the boundary conditions $m = m_0+\ord{e^{-x^2/2}}$ and $\ p = \ord{e^{-x^2/2}}$ is actually well-defined within the effective theory,\footnote{This is similar to the entropy difference defined for the asymptotically flat effective theory, but the coefficient should be tuned up to higher order in $1/n$, so that the boundary flux vanishes. }
\begin{equation}
    \frac{1}{n} \rho_F \equiv \rho_M+\partial_x \phi-\kappa_0 \rho_S,\quad
    \frac{1}{n} j_F \equiv j_M-\partial_t \phi-\kappa_0 j_S,
    %\frac{1}{n} ( \rho_F,j_F) := (\rho_M+\partial_x \phi,j_M-\partial_t \phi)-\kappa_0 (\rho_S,j_S) ,
\end{equation}
where $\kappa_0$ is the boundary temperature~\eqref{eq:bdry-temp}.
By choosing $\phi$ as
\begin{equation}
    \phi = \frac{e^\frac{x^2}{2}}{n(\alpha+1)}(p \log m+ 2 \partial_x m),
\end{equation}
the free energy has vanishing boundary flux\footnote{One can check this by expanding $j_F$ with $m=m_0+\epsilon \delta m,\, p= \epsilon \delta p$ and requiring $\Or{\epsilon}$ vanishes.}
\begin{equation}
    j_F \to \ord{e^{-x^2/2}}. \quad (x\to \pm \infty)
\end{equation}
This condition can be understood as
\beq
\delta M_{\rm bulk} = \kappa_0\delta S_{\rm bulk},
\eeq
which is a natural relation under the heat bath.

In the hydrodynamical expression, the free energy density is given by
\begin{equation}
    \rho_F = \frac{e^{\frac{x^2}{2}}}{\alpha+1}\left[\frac{(\partial_x m)^2}{2m}+\frac{1}{2}mv^2-(\alpha-1)m(\log (m/m_0)-1)\right].
\end{equation}
At $x\to \pm\infty$, the free energy behaves as
\beq
\rho_F = \frac{\alpha-1}{\alpha+1}e^{\frac{x^2}{2}}m_0+\Or{e^{-x^2/2}}.
\eeq
Then we can define the total free energy measured relative to the uniform string $m(t,x)=m_0$ as
\begin{equation}
    \Delta {\cal F}(t) := \int_{-\infty}^\infty (\rho_F-\rho_F|_{\rm uniform})dx.
\end{equation}
The monotonicity automatically follows from that of the entropy current,
\begin{equation}
    \frac{d\Delta {\cal F}}{dt} = -\frac{2}{\alpha+1}\int_{-\infty}^\infty e^\frac{x^2}{2}m (\partial_x v)^2 dx \leq 0
\end{equation}
In particular, for the static solution with $m=e^{\cal R}$, the free energy is obtained by
\begin{equation}
%   \Delta {\cal F} = \int_{-\infty}^\infty \frac{e^{\frac{x^2}{2}}}{\ell^2+1}\left[e^\cR \left(\frac{1}{2}(\cR')^2-(\ell^2-1)(\cR-\cR_0-1)\right)-(\ell^2-1)e^{\cR_0}\right]dx.
    \Delta {\cal F} = \int_{-\infty}^\infty \frac{ e^{\frac{x^2}{2}}}{\alpha+1}\left[e^\cR \left(\frac{1}{2}(\cR')^2-(\alpha-1)(\cR-\cR_0-1)\right)-(\alpha-1)e^{\cR_0}\right]dx.
    \label{eq:freeF-static}
\end{equation}
Note that to obtain the free energy up to a certain order of $1/n$, one needs know one higher order of the mass, entropy and temperature.% In the auxiliary file, we present the mass, entropy and temperature up to NNNLO and the free energy up to NNLO.

\paragraph{Variation of free energy}
We can exploit some useful information on the dynamics from the free energy.
Let us consider a small variation from a static configuration,
\beq
m = e^{\cR(x)} (1+\varepsilon\delta m(t,x)),\quad v = \varepsilon\delta v(t,x),
\eeq
the first order variation of the free energy becomes
\beq
\delta^{[1]}\Delta F  = -\int_{-\infty}^\infty \frac{e^{x^2/2}e^\cR}{\alpha+1}
\left(\cR''+\frac{1}{2}(\cR')^2+x\cR'+(\alpha-1)(\cR-\cR_0)\right)\delta m\, dx,
\eeq
where we assume the variation decays at $x\to \pm\infty$ as $\delta m,\delta v =\Or{e^{-x^2/2}}$.
Therefore, the first order variation vanishes if and only if $\cR(x)$ is the solution of the static equation~\eqref{eq:staticeq}. This means that the static free energy~\eqref{eq:freeF-static} serves as an effective action for the static equation.
The similar action for the static effective equation was proposed in~\cite{Dandekar:2017aiv}. They also pointed that the action is proportional to $E-T_0 S$, which is equivalent to our free energy.

The second order variation provides information on the linear stability around the static solution.
Expanding the free energy up to $\Or{\varepsilon^2}$, we obtain
\beq
\delta^{[2]} \Delta F  = \int_{-\infty}^\infty \frac{ e^{x^2/2}e^\cR}{2(\alpha+1)}\left((\partial_x \delta m)^2
-(\alpha-1)\delta m^2 + \delta v^2\right)\, dx.\label{eq:freeF-2ndvar}
\eeq
Since the free energy cannot increase in time, the negativity of this integration indicates the solution $\cR(x)$ is unstable to the given onset of the fluctuation $\delta m(t,x), \delta v(t,x)$ at time $t$. The variation $\delta v$ gives positive definite contribution, and hence does not involve the instability. Since it is difficult to show the instability for a general onset $\delta m$, we assume $\delta m$ can be expanded in terms of the orthogonal functions,
\beq\label{eq:dm-expansion}
\delta m(t,x) =\sum_{i=0}^\infty a_i(t)H_i(x/\sqrt{2})e^{-x^2/2}.
\eeq
Plugging this into eq.~\eqref{eq:freeF-2ndvar} with $\delta v=0$, we obtain
\beq
\delta^{[2]} \Delta F  = \sum_{i,j=0}^\infty {\cal H}_{ij} a_i a_j ,\label{eq:dfreeF2-Hij}
\eeq
where we introduced the Hesse matrix
\beq
{\cal H}_{ij} :=\int_{-\infty}^\infty \frac{e^{-x^2/2}e^\cR}{4(\alpha+1)}\left(H_{i+1} \left(x/\sqrt{2}\right)H_{j+1}\left(x/\sqrt{2}\right)-2(\alpha-1)H_i\left(x/\sqrt{2}\right) H_j\left(x/\sqrt{2}\right) \right)\,dx.
\label{eq:def-Hij}
\eeq
If ${\cal H}_{ij}$ is not a positive definite matrix, we can prove the existence of an instability. However, it is not practical to evaluate an infinte-dimensional matrix if $\cR(x)$ is constructed only numerically. Instead, we expect only a few long wavelength modes to be involved to the instability, and thus we can often truncate the expansion at finite order to make ${\cal H}_{ij}$ finite.

For the uniform funnels $\cR(x) = \cR_0$, eq.~\eqref{eq:dfreeF2-Hij} is easily evaluated as
\begin{equation}
    \delta^{[2]} \Delta F = \frac{\sqrt{2\pi}e^{\cR_0}}{2(\alpha+1)}\sum_{k=0}^\infty 2^k k! (k+2-\alpha)a_k^2.
\end{equation}
This indicates each mode becomes an unstable onset if $\alpha > k+2$. One can see that for $\signK=-1$ and $\signK=0$, the uniform string is always stable. In the $\signK=1$ case, we have a threshold radii of $r_0 = 1/\sqrt{k+2}$ for each onset, which in turn shows the stability of fatter strings with $r_0>1/\sqrt{2}$. This is consistent with the perturbative analysis we perform later in section~\ref{sec:uniformperts}.  The stability of the uniform strings can also be shown for general onset $\delta m$ by using a Wirtinger-type inequality\footnote{
Suppose that $\phi(t,x):=e^{x^2/2}\delta m(t,x)$ is a smooth function of $x \in (-\infty,\infty)$ which have at most polynomial growth for $|x|\to\infty$.
Then, a simple calculation follows
\beq
  \int_{-\infty}^\infty e^{x^2/2}((\partial_x \delta m)^2-\delta m^2)dx = \int^{\infty}_{-\infty} e^{-x^2/2} (\partial_x \phi)^2 dx \geq 0.
\eeq
}
\beq
    \delta^{[2]} \Delta F \geq    \frac{e^{\cR_0}}{2(\alpha+1)} \int_{-\infty}^\infty e^{x^2/2} \left( (2-\alpha)(\delta m)^2+\delta v^2 \right)dx
\eeq
which again guarantees the linear stability of the uniform funnels for $r_0>1/\sqrt{2}$.
%
%%%%%%%%%%

%--------------------------------------------------------------
\section{Thermodynamics and stability of exact static solutions}\label{sec:thermostabexact}
%--------------------------------------------------------------

In this section, we study the thermodynamics and stability of the static solutions.  But first, we briefly explain some terminology commonly used in holographic setups with boundary black holes \cite{Hubeny:2009ru}.  When horizons of boundary black holes extend into bulk horizons, the resulting bulk horizons can either be connected or disconnected.  Connected configurations are called ``funnels," which disconnected configurations are called ``droplets."

In the current setup, we only have a single connected horizon, so technically all configurations are funnels.  However, we also have situations where the mass function exhibits Gaussian decay.  It is natural to expect that these connections are severed at finite $D$, and so we treat exponentially small mass functions as equivalent to the absence of a horizon.

We therefore have two choices of boundary behaviors on each side: approach to the uniform solution $m\to m_0$ or the Gaussian decay $m\to0$. We will look at primarily three possible configurations according to their boundary choice: (1) Gaussian blobs $m(\pm \infty)=0$, which resemble bulk black holes that are fully disconnected from the boundary (2) Funnels or double droplets with $m(\pm \infty)=m_0$, (3) Single droplets with $m(-\infty)=0$ and $m(\infty)=m_0$, and their physically equivalent versions with $x\to-x$.  There are more complicated configurations such as double droplets with Gaussian blobs between them, but we will not address them here.

Using this terminology, the uniform strings we have described fall under case (2) and are equivalently called uniform funnels. We will therefore use the terms ``uniform string" and ``uniform funnels" interchangably.

\subsection{Gaussian blobs}
\label{sec:GaussianBlob}

The static equation~(\ref{eq:staticeq-m-3rd}) has a simple exact static solution
\beq\label{eq:gblob}
m(x)=m_0\,e^{-\frac{\alpha+1}{2} x^2}\,.
\eeq
%Since $v=0$ we have
%\beq
%p=\partial_x m= -(r_0^{-2}+1) x\, m(x)\,.
%\eeq
In the rescaled coordinate~\eqref{eq:coordinate-AF},
the limit $r_0\to 0$ recovers the usual static blob.

\paragraph{Gaussian blob as large-$D$ Schwarzschild-AdS black hole.}

The solution \eqref{eq:gblob} can actually be recovered as the large $D$ limit of the Schwarzschild-AdS solution%, which in Eddington-Finkelstein coordinates can be written
\beq
%ds^2 =L^2\lp -F(R) dt^2+2 dt dR +R^2(d\theta^2+\cos^2\theta d\Omega_{n+1})\rp\,.
ds^2 =L^2\lp -F(R) dt^2+\frac{dR^2}{F(R)} +R^2(d\theta^2+C_K^2(\theta) d\Sigma_{n+1})\rp\,
\eeq
with
\beq
 F(R) = \signK + R^2 - \frac{(\signK+r_0^2)r_0^{n+1}}{R^{n+1}},
\eeq
where the horizon is present at $R=r_0$. First, we change to the coordinate fit to the AdS black string background~(\ref{eq:AdS-blackstring-sol})
\beq
  R = \sqrt{(\signK+r^2)\sec^2 z-\signK},\quad T_K(\theta) = \frac{\sin z}{r}.
\eeq
Since the horizon exists around $(r,z)=(r_0,0)$, the large $n$ limit with the large $D$ near horizon coordinate for AdS strings~(\ref{eq:coordtransf}) keeps the horizon
\beq
   F(R) \simeq r_0^2(1+\alpha) \left(1 -\frac{e^{-\frac{1+\alpha}{2}x^2}}{e^\rho}\right).
\eeq
By rescaling the time coordinate by $t\to t/r_0$ and switching to the Eddington-Finkelstein coordinate, we can verify that to leading order in $1/n$ the metric becomes of the same form as \eqref{eq:metans}, \eqref{eq:ACHu} for the Gaussian solution \eqref{eq:gblob}.

\paragraph{Bouncing blob}
\label{sec:bouncingBlob}
The boost transformation~\eqref{eq:boostsym} on the Gaussian blob solution leads to the following stationary solution
\beq
    m(t,x) = m_0 e^{-\frac{x^2}{2}} e^{-\frac{\alpha}{2}(x+V(t))^2},\quad v(x) = -V'(t).
\eeq
For $\signK=1$, this gives a Gaussian blob with a oscillatory motion. This corresponds to the AdS black holes bouncing back and forth in the AdS interior.

\subsubsection{Stability of AdS-black holes/ Gaussian blobs}\label{sec:blobperts}
We study the dynamical perturbation of the Gaussian blob presented in section \ref{sec:GaussianBlob}. Perturbing the solution that corresponds to an AdS-black hole according to
\begin{align}
    m(t,x) &= \,e^{-\frac{\alpha+1}{2} x^2} \left(m_0 + e^{\Omega t}   \delta m(x)\right)\,,\\
    v(t,x) &= e^{\Omega t} \delta v(x)\,.
\end{align}
the equations of motion (\ref{eq:dtm}) and (\ref{eq:dtmv}) yield the perturbation equations
\begin{align}
&- \alpha x \delta v+\delta v'+\Omega  \delta m = 0\label{eq:blob-dtm}\\
&\left(\alpha^2 x^2+\alpha \left(x^2-1\right)+\Omega \right) \delta v -x \delta v' -\delta v'' \nn\\
   -&\left(\alpha+1\right) x \Omega  \,\delta m +(1+\Omega)\,  \delta m' +\alpha x\, \delta m'' -\delta m^{(3)}  =0\,.
\end{align}
These equations can be written as a fourth order master equation
\begin{align}
0=\Omega  \left(\Omega -2 \alpha\right) \delta m+ \alpha x (1+2  \Omega ) \delta m' +\left( \alpha^2 x^2- \alpha-2  \Omega -1\right) \delta m'' 
 -2  \alpha x \delta m^{(3)} +\delta m^{(4)}.
 \label{eq:masterBlobPert}
\end{align}
We find that this is equivalent to the product of two commuting second order differential operators
\begin{align}
    \mathcal{D}_1\mathcal{D}_2\delta m = \mathcal{D}_2 \mathcal{D}_1 \delta m=0,
\end{align}
where
\begin{align}
    \mathcal{D}_i := \partial_x^2 - \alpha x \partial_x+ \alpha \lambda_i
    \label{eq:DiffOpBlobPert}
\end{align}
and $\lambda_i$'s are the roots of 
\begin{align}
\alpha^2\lambda^2 + \alpha(2\Omega-\alpha+1)\lambda + \Omega(\Omega-2\alpha)=0.
\label{eq:blob-disperse-cond}
\end{align}
%This constraint will determine the quasi-normal mode frequencies $\omega$.
The master equation~\eqref{eq:masterBlobPert} admits normalizable solutions
\begin{align}
    \delta m(x) = H_k \left( \sqrt{\frac{|\alpha|}{2}} x\right),\quad k=0,1,2,\dots
\end{align}
only if $\lambda = k$ is a solution of eq.~\eqref{eq:blob-disperse-cond}.
From eq.~\eqref{eq:blob-dtm}, $\delta v(x)$ is also obtained as
\begin{align}
    \delta v(x) = - \frac{2\Omega}{\sqrt{|\alpha|}} H_{k-1} \left( \sqrt{\frac{|\alpha|}{2}} x\right).
\end{align}
for $k \geq 1$. $k = 0$ leads to non-normalizable mode in $\delta v(x)$ and hence not allowed.
By solving eq.~\eqref{eq:blob-disperse-cond} for $\lambda=k=1,2,3,\dots$, the growth rate is given by
\begin{align}
    \Omega = - (k -1)\alpha \pm i \sqrt{\alpha k +\alpha^2(k -1)}\, .
\end{align}
We conclude that all normalizable modes are stable, as expected for AdS-black holes and there is a purely oscillatory mode for $k = 1$, that corresponds to the onset of the 'bouncing blob' described in section \ref{sec:bouncingBlob}.

%\DL{There are possibly further modes which have exponential fall-off in the $m$ variable that I haven't yet described here.}
%
%%
%\subsection{Funnels}
%TODO
%\begin{itemize}
%    \item Plots of phases and shapes. An over view of the result, phase and stability of funnels, combined with numerics and perturbations
%    \item Numerical scheme
%\end{itemize}

%\subsection{Perturbative analysis of near neck effective theory: zero modes, stabilities and non-uniform funnels}

%%
%
\subsubsection{Gaussian blobs as fat funnels and black tsunamis}\label{sec:blobandtsunami}

Though we have managed to recover these Gaussian blobs from the large $D$ limit of the Schwarzschild-AdS solution, they can also be interpreted as ``fat" funnels, and are often the preferred endpoint of the large-D Gregory-Laflamme instability in this system, as was show in \cite{Emparan:2021ewh}.  Here, we briefly review the the results in \cite{Emparan:2021ewh} and explain this interpretation.

It was found, by numerical time evolution, that unstable uniform black strings sometimes develop a large bulge which grows without bound.  The endpoint thus seems to be a solution which lies outside of the large-$D$ effective equations.  However, this endpoint can be interpreted as a Gaussian blob in the following way.  Consider the the mass function $m(t,x)$ and rescale it via \eqref{eq:scaling} as $m_s(t,x)=m(t,x)/m(t,0)$. Then even though $m(t,0)$ grows without bound as $t\to\infty$, $m_s(t,x)$ approaches the Gaussian blob.  So the final endstate resembles an infinitely rescaled Gaussian solution. 

For the Gaussian blob, its free energy, relative to the uniform solution, can be formally written as
\begin{align}
    \Delta {\cal F} = ({\rm finite \ terms}) + \frac{1-\alpha}{1+\alpha}m_0 \int_{-\infty}^\infty  dx\,e^\frac{x^2}{2},
\end{align}
which diverges since the boundary values are different than for the string. But we can still extract relevant information from this formal expression, since the overall sign of $\Delta {\cal F}$ changes with $\alpha$.  In particular, when $\alpha>1$ (that is, when $\signK=1$ and $r_0<1$), the Gaussian blob has negative infinite free energy compared to any string, uniform or non-uniform. Hence, the Gaussian solution is dominant over thin strings.  The time-dependent simulations in \cite{Emparan:2021ewh} are fully consistent with this thermodynamic analysis. 

\subsection{Stability of uniform funnels}\label{sec:uniformperts}
Like black strings in flat space, uniform funnels exhibit a Gregory-Laflamme instability when its thickness becomes thin enough. Here we show the appearance of the instability using linear perturbation theory.  Let us consider the following perturbation,
\begin{equation}
 m(t,x) = 1 +  e^{\Omega t} \delta m(x),\quad p(t,x) =  e^{\Omega t} \delta p(x),
\end{equation}
where the mode functions are assumed to have Gaussian decay at $x\to\pm\infty$ so that the boundary behavior remains the same.
The linearized equation becomes
\begin{align}
& \delta m''+x\, \delta m'+\delta p'+x \,\delta p-\Omega\, \delta m = 0,\\
& \delta p'' + x \,\delta p' -(1+\Omega) \,\delta p-(\alpha+1)\, \delta m' = 0.
\end{align}
Eliminating $\delta p$, we find a fourth order master equation
\begin{align}
  \delta m^{(4)}+2 x \delta m^{(3)}+ (\alpha+x^2-2 \Omega+1)\delta m''
   + x \left(\alpha-2 \Omega \right)\delta m'+\Omega(\Omega +2)\delta m=0.
\end{align}
As in the previous section, the master equation is again equivalent to the product of two commuting differential operators
\begin{align}
{\cal D}_1 {\cal D}_2 \delta m = {\cal D}_2 {\cal D}_1 \delta m = 0,
\end{align}
where ${\cal D}_i$ are given by
\begin{align}
    {\cal D}_i :=\partial_x^2 + x \partial_x + \lambda_i
\end{align}
and $\lambda_i$'s are the roots of
\begin{align}
    \lambda^2+(2\Omega-\alpha+1)\lambda+\Omega(\Omega+2)=0.\label{eq:ubspert-cond}
\end{align}
The normalizable modes exist if
\begin{align}
    \lambda=k+1 \quad (k=0,1,2,\dots)
\end{align}
is a root of eq.~\eqref{eq:ubspert-cond}, where the mode function is given by
\begin{align}
    \delta m = e^{-x^2/2} H_{k}(x/\sqrt{2}).
\end{align}
From eq.~\eqref{eq:ubspert-cond}, 
the dispersion relation is obtained as
\begin{align}
    \Omega = - 2-k\pm \sqrt{2+k+(k+1)\alpha}.
\end{align}
It is easy to see that the zero modes exists for
\begin{align}
    \alpha = k+2.\label{eq:zero-mode-lin}
\end{align}
Though we do not give the technical details, we also solved the perturbation equation up to ${\cal O}(1/n^2)$ for the first unstable mode ($k=0$) which gives the dispersion
\begin{align}
    & \Omega =-2+ \sqrt{\alpha+2}+
    \fr{n}\left(\frac{(7+\alpha) \sqrt{\alpha+2}}{2 +\alpha}-6\right)\nonum
&   +\fr{n^2}\left(\frac{2 \left(\pi ^2 -1-\alpha\right)}{\alpha+1}+\frac{\sqrt{\alpha+2}(-32 \pi ^2 +57-(24 \pi ^2-105)\alpha-(4 \pi ^2-39) \alpha^2-9\alpha^3)}{6(1+\alpha)(2+\alpha)^2}\right).
\end{align}
The zero mode occurs for
\begin{align}
    \alpha = 2 \left(1+\frac{3}{n}+\frac{1}{n^2}\right).\label{eq:firstzeromode-threshold}
\end{align}
%\begin{align}
%    r_0 = \fr{\sqrt{2}}\left(1 - \frac{3}{2 n} + \frac{23}{8 n^2}\right).
%\end{align}
One can check this is consistent with the zero mode analysis up to $1/n$ in \cite{Marolf:2019wkz}.

\subsubsection{Weakly non-uniform funnels}\label{sec:pertnonuniform}

Like black strings in flat space, we also have several non-uniform families originating from zero modes~(\ref{eq:zero-mode-lin}). Here we study these non-uniform branches when the deformations are perturbatively small.

We begin with assuming the solution is perturbatively expanded from the uniform funnel
\begin{align}
    \cR(x) =  \varepsilon \delta^{[1]} \cR(x) + \varepsilon^2 \delta^{[2]}\cR(x)+\cdots\;.
\end{align}
As in eq.~\eqref{eq:staticeq-bdry-bhv}, we impose the normalizable condition such that the solution at each order decays as $e^{-x^2/2}$ at $|x|\to\infty$.
Expanding eq.~(\ref{eq:staticeq}) in a small $\varepsilon$, we obtain
\begin{align}
\delta^{[i]} \cR''+ x \, \delta^{[i]} \cR'+(\alpha-1) \delta^{[i]}\cR'={\cal S}^{[i]},
\end{align}
where the source term ${\cal S}^{[i]}$ consists of lower order solutions. Now, let us start from $k$-th zero mode perturbation on the uniform funnel
\begin{equation}
\alpha =2+k,\quad \delta^{[1]} \cR(x) =u_k(x):= H_{k}\left(\frac{x}{\sqrt{2}}\right) e^{-\frac{x^2}{2}},
\end{equation}
where $u_k$ is a solution of the linear equation
\beq
u_k''+x u_k'  + (k+1) u_k =0. \label{eq:uk-eq}
\eeq

At second order, we expect the non-linear perturbation backreact to the quantization condition
\begin{align}
 \alpha = k+2+\alpha_1\, \varepsilon.
\end{align}
Thus, the second order equation becomes
\beq
\delta^{[2]} \cR''+ x  \delta^{[2]} \cR'+(k+1) \delta^{[2]}\cR={\cal S}^{[2]}-\alpha_1 u_k,\label{eq:nubs-secondordeq}
\eeq
where the source term is given by
\beq
  {\cal S}^{[2]} = \fr{2}( (u_k')^2-(k+1)u_k^2).
\eeq
Multiplying by $e^{x^2/2} u_k$ on the left hand side of eq.~\eqref{eq:nubs-secondordeq}, one can see
\begin{align}
e^{x^2/2} u_k \times {\rm l.h.s \ of\ } \eqref{eq:nubs-secondordeq} =  \partial_x \left[e^{x^2/2}\left( u_k   \partial_x \delta^{[2]} \cR-  \delta^{[2]} \cR\partial_x u_k\right) \right], %+ e^{x^2/2} \left( u_k''+x u_k'  + (k+1) u_k   \right)\delta^{[2]} \cR.
\end{align}
where we used eq.~\eqref{eq:uk-eq}.
Assuming $\delta^{[2]} {\cal R}$ decayes as $e^{-x^2/2}$ at $|x|\to \infty$, the integration over $(-\infty,\infty)$ should vanishes, which yields the normalizable condition
\begin{align}
  0=  \int_{-\infty}^\infty e^{x^2/2} \left(\fr{2}u_k (u_k')^2-\frac{k+1}{2} u_k^3 - \alpha_1 u_k^2\right) .
\end{align}
By integrating by part with eq.~\eqref{eq:uk-eq}, we find
\begin{align}
\alpha_1 = -\frac{k+1}{4} \frac{\int^\infty_{-\infty}  e^{x^2/2}u_k(x)^3 dx}{\int_{-\infty}^\infty e^{x^2/2} u_k(x)^2 dx} = -\frac{k+1}{\sqrt{\pi}\, 2^{k+2}\, k!} \int^\infty_{-\infty}  e^{-2u^2}H_k(u)^3 du.\label{eq:res-alpha1}
%\mu_{p} = \fr{\sqrt{\pi} 2^{p+1} p!} \int^\infty_{-\infty} du e^{-2u^2} \left(2p^2H_{p-1}(u)^2H_p(u)-4puH_{p-1}(u)H_p(u)^2+(1+p-2u^2)H_p(u)^3\right).
\end{align}
Using the parity of the Hermite polynomials, one can show odd modes do not get corrections at this order
\begin{align}
\alpha_1\bigr|_{k:{\rm odd}} = 0,
\end{align}
which means the correction for odd modes begins at $\ord{\varepsilon^2}$. This is expected since the odd mode perturbation gives physically identical deformations for either sign of $\varepsilon$, and hence the branching should not be two sided.

For the even modes, the perturbations are even about $x$ and depend on the sign of $\varepsilon$.  We call these branches {\it bulges} (and {\it ditches}) where the non-uniform perturbation increases (decreases) the mass function at $x=0$.

The values of the coefficient for the lowest few modes are
\begin{equation}
\alpha_1\bigr|_{k=0,2,4} = - \fr{4\sqrt{2}},\quad -\frac{3}{32\sqrt{2}},\quad  -\frac{45}{512\sqrt{2}}.
\end{equation}
The negative value of $\alpha \bigr|_{k=0}$ indicates that, from the first zero mode, the bulge branch appears in the fatter side $r_0>1/\sqrt{2}$ where the uniform funnel becomes stable, while the ditch branch appears in the thinner, unstable side.%\footnote{For higher modes, one should bear in mind that $H_{2k}(0)$ has flipping signature about the parity of $k$.}

\paragraph{Free energy}
With the second order perturbation, we can now evaluate the free energy of each branch. By integrating by parts, we obtain
\beq
 \Delta {\cal F} = -\frac{\varepsilon^3}{2(k+3)}\left(\alpha_1 \int_{-\infty}^\infty e^{x^2/2} (\delta^{[1]} \cR)^2dx+\frac{k+1}{6}\int_{-\infty}^\infty e^{x^2/2} (\delta^{[1]} \cR)^3 dx\right)=- \frac{\sqrt{2\pi} 2^k k! \alpha_1}{6(k+3)}\varepsilon^3.
\eeq
And for the lowest modes, we have
\beq
  \left.\Delta {\cal F}\right|_{k=0,2,4} = \frac{\sqrt{\pi}}{72}\varepsilon^3,\quad \frac{\sqrt{\pi}}{40}\varepsilon^3,\quad \frac{45\sqrt{\pi}}{56}\varepsilon^3.
  \label{eq:freeen-ditches}
\eeq
For the first branch, one can see that the ditch branch $(\varepsilon<0)$ has lower free energy and bulge branch $(\varepsilon>0)$ has larger free energy than the uniform string.   So long as the deformations are small, it is therefore thermodynamically possible for unstable uniform strings to evolve into non-uniform ditch configurations, but not bulge configurations.

\paragraph{Critical dimension in the first deformed branch}
For the first branch, by repeating the same analysis up to ${\cal O}(n^{-2})$, we obtain the solution expanded both in $\varepsilon$ and $1/n$,
\begin{equation}
 \cR(x) = \varepsilon \left(\delta^{[1,0]} \cR(x)+\frac{\delta^{[1,1]} \cR(x)}{n}+\frac{\delta^{[1,2]} \cR(x)}{n^2}\right)+\varepsilon^2 \left(\delta^{[2,0]} \cR(x)+\frac{\delta^{[2,1]} \cR(x)}{n}+\frac{\delta^{[2,2]} \cR(x)}{n^2}\right)
\end{equation}
with
\begin{equation}
 \frac{\alpha}{\alpha_{\rm GL}} = 1 - \frac{\varepsilon}{8\sqrt{2}} \left(1-\frac{53}{8n} -\frac{5823+64\pi^2}{1152 n^2} \right),
\end{equation}
where $\alpha_{\rm GL}$ is the threshold parameter for the first zero mode~(\ref{eq:firstzeromode-threshold}).
One can see the phase of bulges ($\varepsilon>0$) and ditches ($\varepsilon<0$) flip the side across a certain value of $n$,
\begin{equation}
 n_{\rm crit} = 7.38\dots,\quad {\rm or}\quad D_{\rm crit} = 11.38\dots. \label{eq:static-criticalD}
\end{equation}

\subsubsection{Stability of weakly non-uniform funnels}\label{sec:pertnonuniformstab}
The stability of deformed branches can be studied in the same way as in the uniform black string,
\begin{equation}
 m(t,x) = e^{\cR(x)} (1 + \, e^{\Omega  t} \delta m(x)),\quad p(t,x) =e^{\cR(x)} (\cR'(x)+ \, e^{\Omega t} \delta p(x)),
\end{equation}
where $\cR(x)$ is the static solutions solved by the perturbative expansion in the previous section
\begin{equation}
 \cR(x) =  \varepsilon \, u_k(x)+\varepsilon^2 \,\delta^{[2]} \cR(x)+\dots
\end{equation}
with the parameter
\begin{equation}
\alpha = k+2 + \alpha_1 \,  \varepsilon.
\end{equation}
Under these background, the mode function and growth rate is solved by the perturbative expansion in $\varepsilon$,
\begin{align}
 \delta m(x) = u_k + \delta^{[1]} m(x) \varepsilon+\cdots,\quad
 \delta p(x)=  u_k' +  \delta^{[1]} p(x) \varepsilon+\cdots,
\end{align}
and
\begin{equation}
\Omega = 0 + \Omega_1 \varepsilon + \cdots  ,
\end{equation}
where the zeroth order solution is set that of the static zero mode, since the background solution reduces to the uniform funnel at $\varepsilon=0$. The equation at the first order can be written in the form of master equation with source terms
\begin{align}
    &{\cal D}_1 {\cal D}_2 \delta^{[1]} m =     {\cal D}_2 {\cal D}_1 \delta m^{[1]} \nn\\
    &= 
    - 2(1+k)^2 u_k^2 + 2(2+k-x^2) (u_k')^2+ ((k+1)\alpha_1-2(k+2)\Omega_1) u_k - 4 (k+1)x u_k u_k' 
\end{align}
where the differential operators ${\cal D}_i$ are given by
\begin{align}
    {\cal D}_1 = \partial_x^2 + x \partial_x + k+1,\quad {\cal D}_2 = \partial_x^2 + x \partial_x.
\end{align}
It turns out the operator ${\cal D}_2$ can be integrated as
\begin{align}
    {\cal D}_1 \delta^{[1]}m = \left(\frac{2(k+2)\Omega_1}{k+1}-\alpha_1\right)u_k - (u_k')^2
\end{align}
where we used the property of the Hermite polynomials.
With the same argument as in the previous section, integrating over $(-\infty,\infty)$ after multiplying by $e^{x^2/2}u_k$ yields the normalizable condition for $\delta^{[1]} m$
\begin{equation}
0=\int_{-\infty}^\infty e^{x^2/2}\left[\left(\frac{2(k+2)\Omega_1}{k+1}-\alpha_1\right)u_k^2 - u_k (u_k')^2\right]dx.
\end{equation}
Integrating by part, one can simplify as
\begin{align}
0 =
\left(\frac{2(k+2)}{k+1}\Omega_1 - \alpha_1 \right) \int ^\infty_{-\infty}e^\frac{x^2}{2} u_k^2 dx
- \frac{k+1}{2} \int ^\infty_{-\infty}e^\frac{x^2}{2} u_k^3 dx.
\end{align}
Thus, using eq.~\eqref{eq:res-alpha1}, we obtain
\begin{equation}
 \Omega_1 = - \frac{k+1}{2(k+2)}\alpha_1.
\end{equation}
Particularly, for the first branch, we have
\begin{align}
\Omega_1 = \fr{16\sqrt{2}}.
\end{align}
This shows the bulge deformations unstable and ditches stable, which
is consistent with the thermodynamic consideration~\eqref{eq:freeen-ditches}.

\paragraph{$1/n$ correction to the stability of the first deformed branch}
As in the static deformation, one can continue the similar analysis both with $1/n$ and $\varepsilon$ expansions.
Here we show the result for the first branch
\begin{equation}
\Omega = \frac{\varepsilon }{16 \sqrt{2}}\left(1-\frac{37}{8
   n}-\frac{7797+64 \pi ^2}{384 n^2}\right)
\end{equation}
Thus, we observe the critical dimension in the stability which is consistent with that in the phase diagram~(\ref{eq:static-criticalD}) within the accuracy of ${\cal O}(1/n)$,
\begin{equation}\label{eq:stabcriticalD}
 n_{\rm crit} = 7.54\dots,\quad {\rm or} \quad D_{\rm crit} = 11.54\dots.
\end{equation}

%--------------------------------------------------------------
\section{Numerical static phases and phase diagram}\label{sec:numericalphase}
%--------------------------------------------------------------

We now present numerical static solutions of the effective equation. As before, there are three types of static solutions which correspond to the different boundary conditions, now with $m_0=1$ to fix rescaling symmetry:
\begin{enumerate}
    \item Funnels and double droplets : $m(x\to-\infty)=1,\ m(x\to\infty)=1$
    \item Single droplets : $m(x\to-\infty)=0,\ m(x\to\infty)=1$ (or $m(x\to-\infty)=1,\ m(x\to\infty)=0$)
    \item Gaussian Blobs : $m(x\to -\infty)=0,\ m(x\to\infty)=0$
\end{enumerate}
Among the three, the third family only admits the Gaussian blobs shown earlier in sec\ref{sec:GaussianBlob}, while the other two admit a variety of deformed branches.

\subsection{Funnels}\label{sec:funnels}
\begin{figure}
    \centering
    \includegraphics[width=11cm]{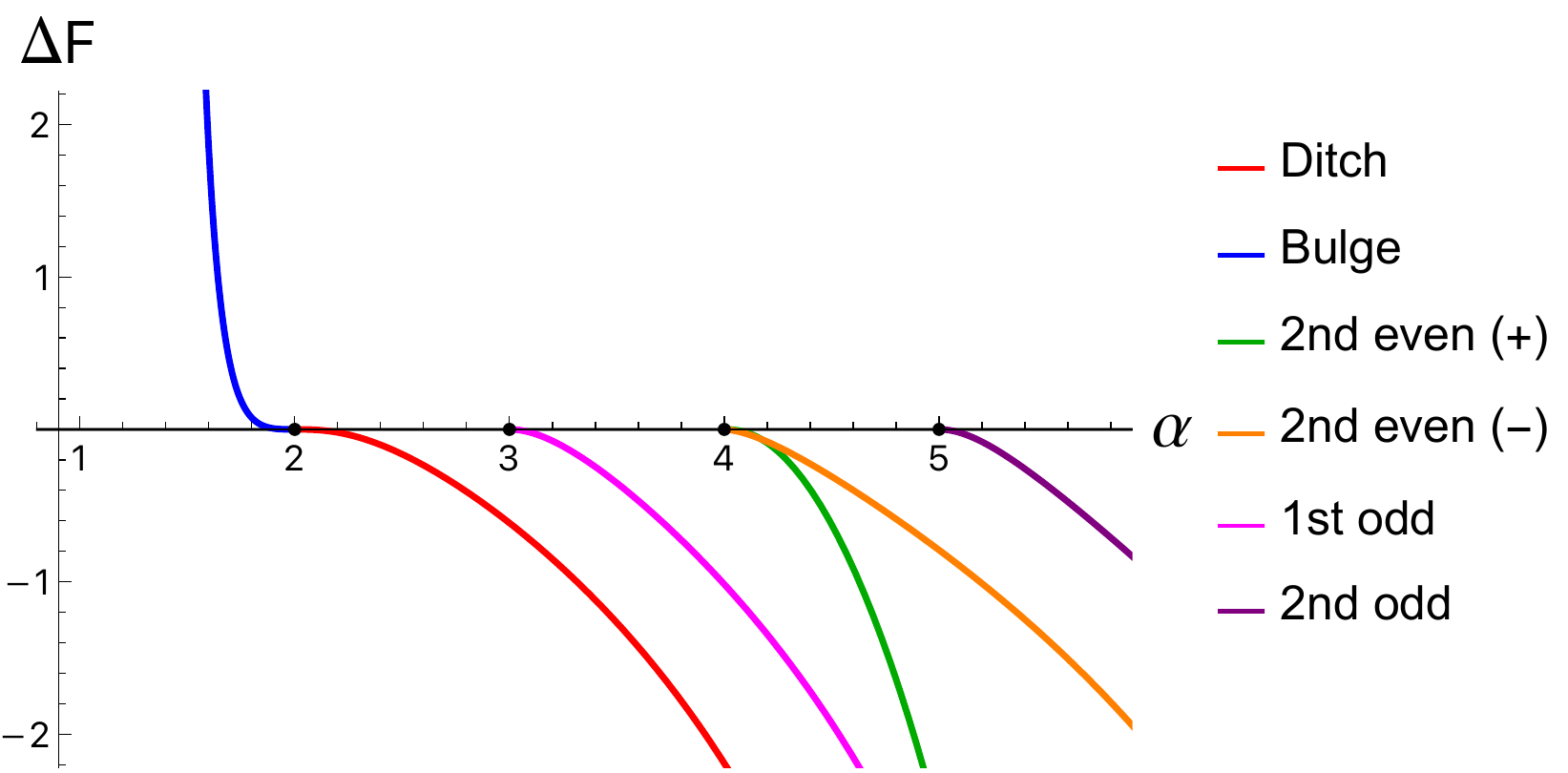}
    \caption{Free energy of the deformed funnels compared with the uniform funnel.
    The dots denotes the onsets of instability on the uniform funnel. For the second even branches, $(\pm)$ denotes the signature of $m(0)$.}
    \label{fig:phasediagram-funnels}
\end{figure}

\begin{figure}
    \centering
    \includegraphics[width=7.1cm]{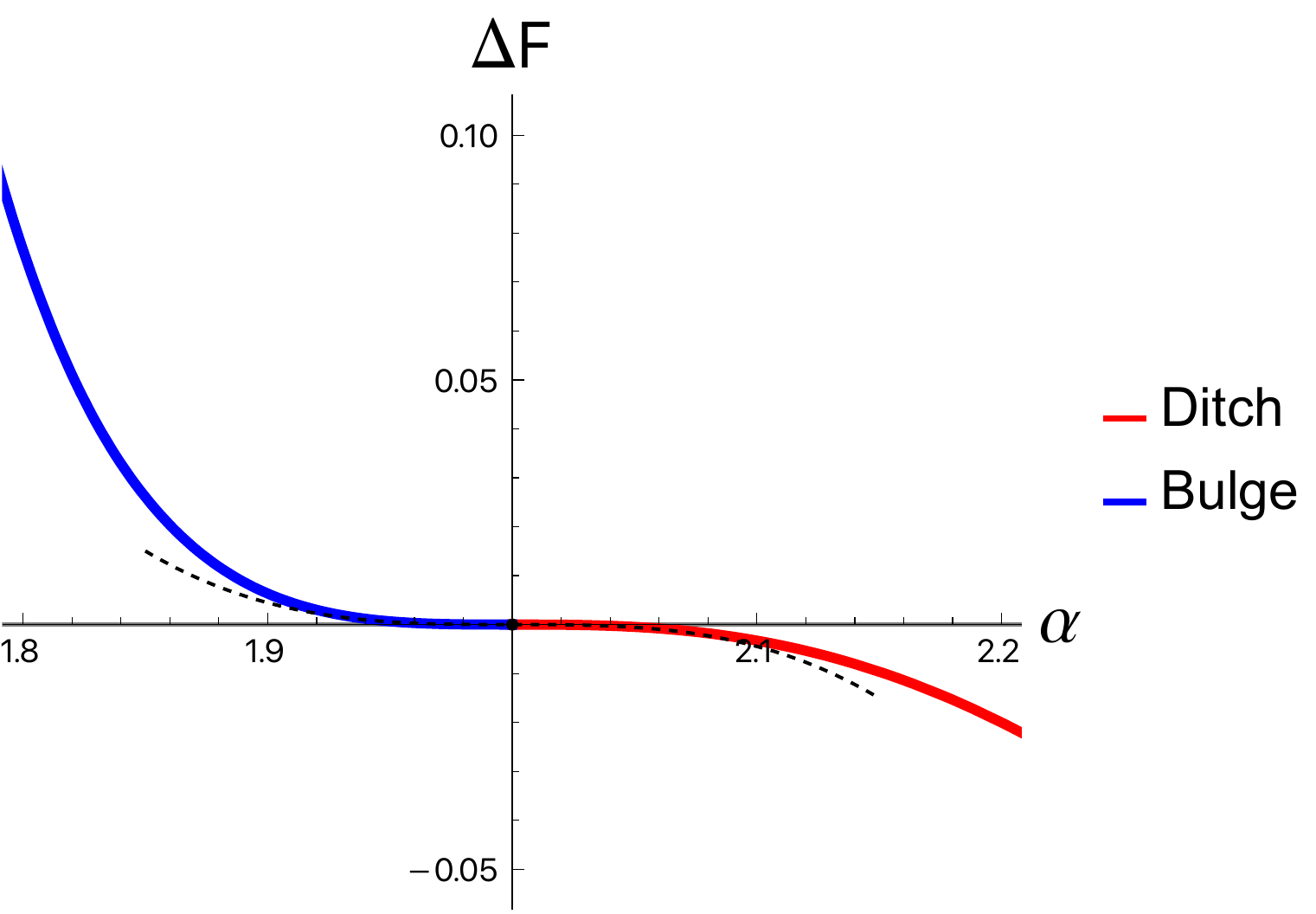}
    \includegraphics[width=7.8cm]{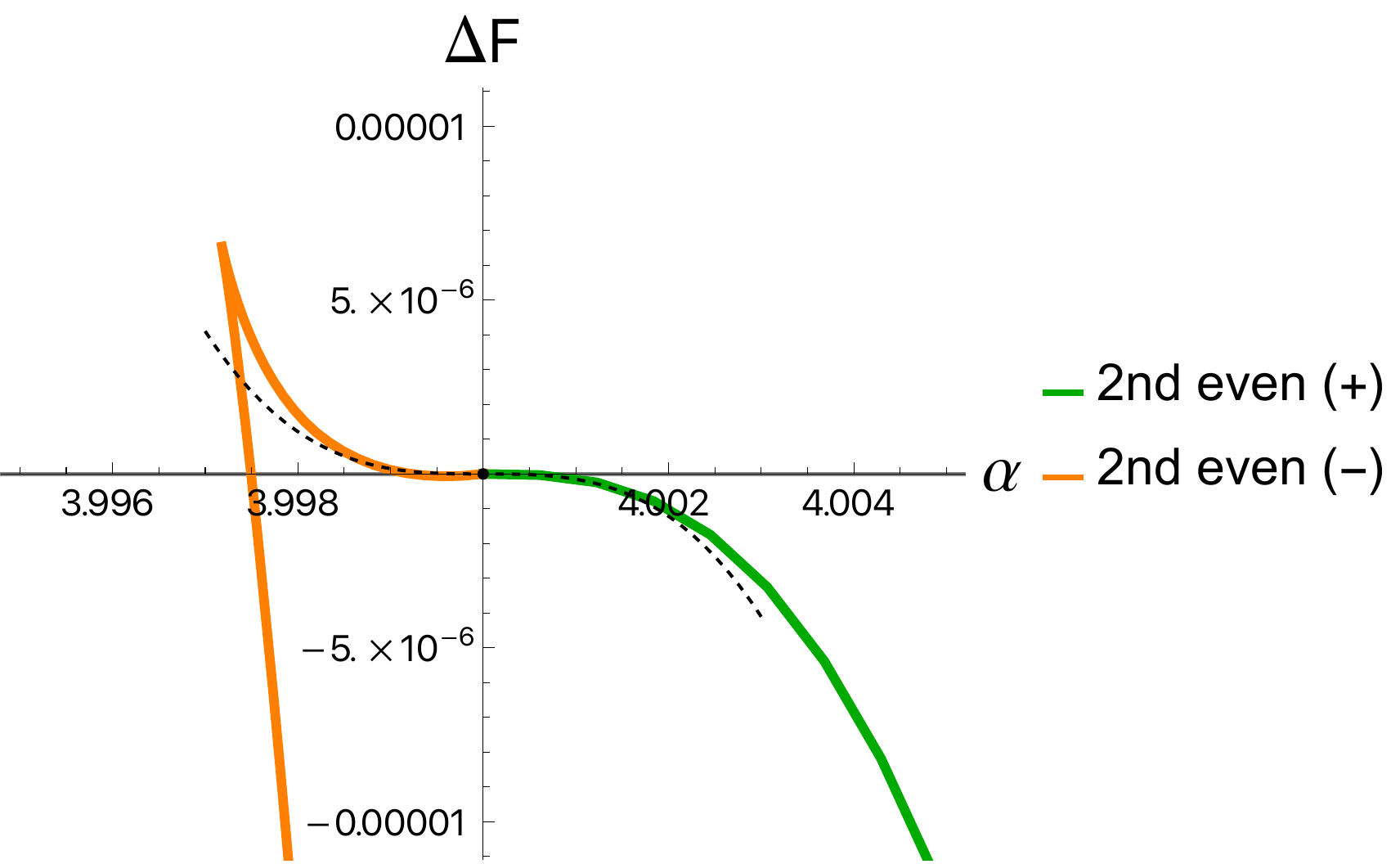}
    \caption{The comparison with the perturbative result (dashed curves) near the branching points of the first and second even modes. The second branch admits a tiny cups in the (-) side.}
    \label{fig:phasediagram-funnels-enlarge}
\end{figure}

\begin{figure}
    \centering
    \includegraphics[width=7cm]{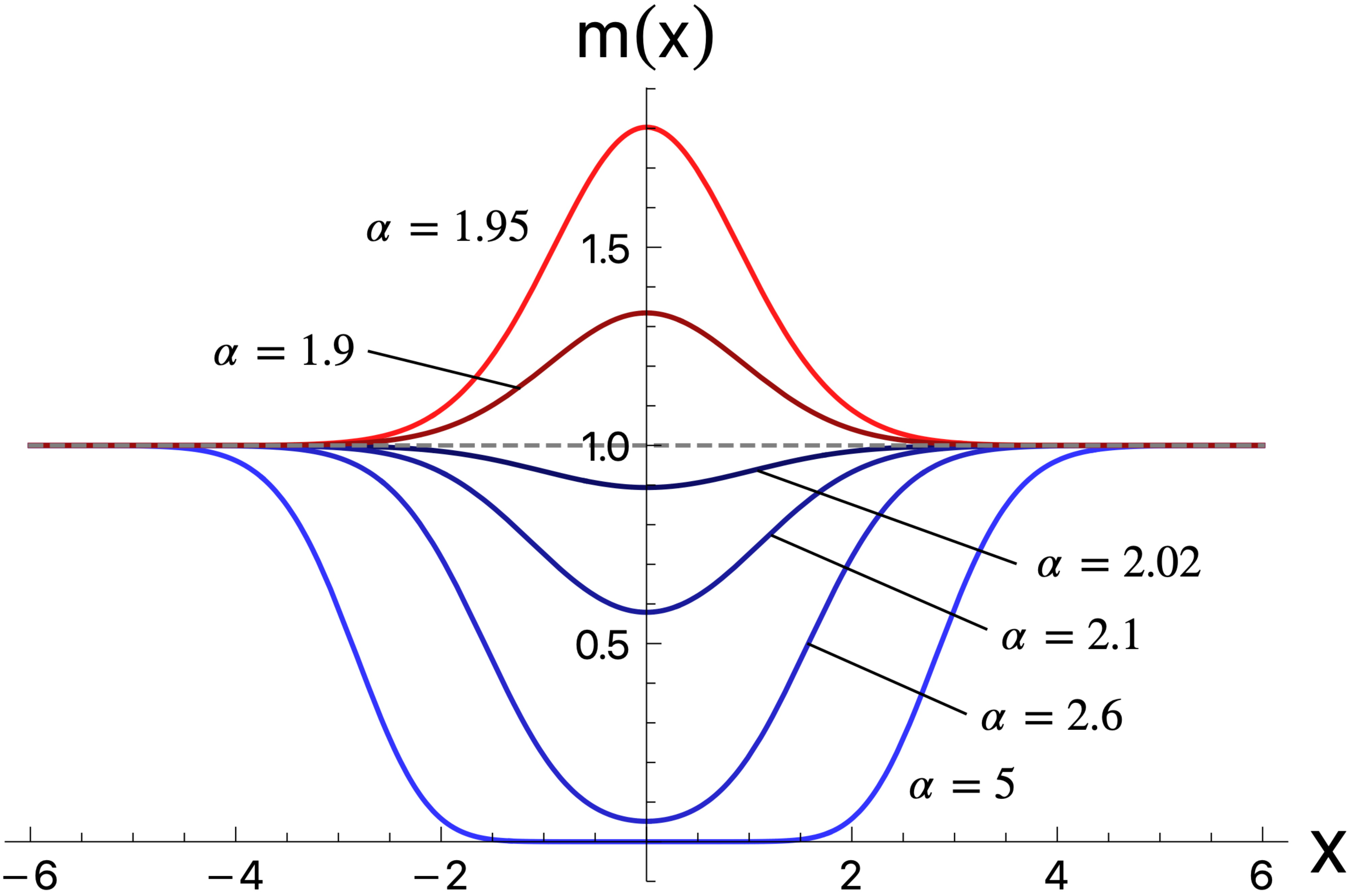}
    \includegraphics[width=7cm]{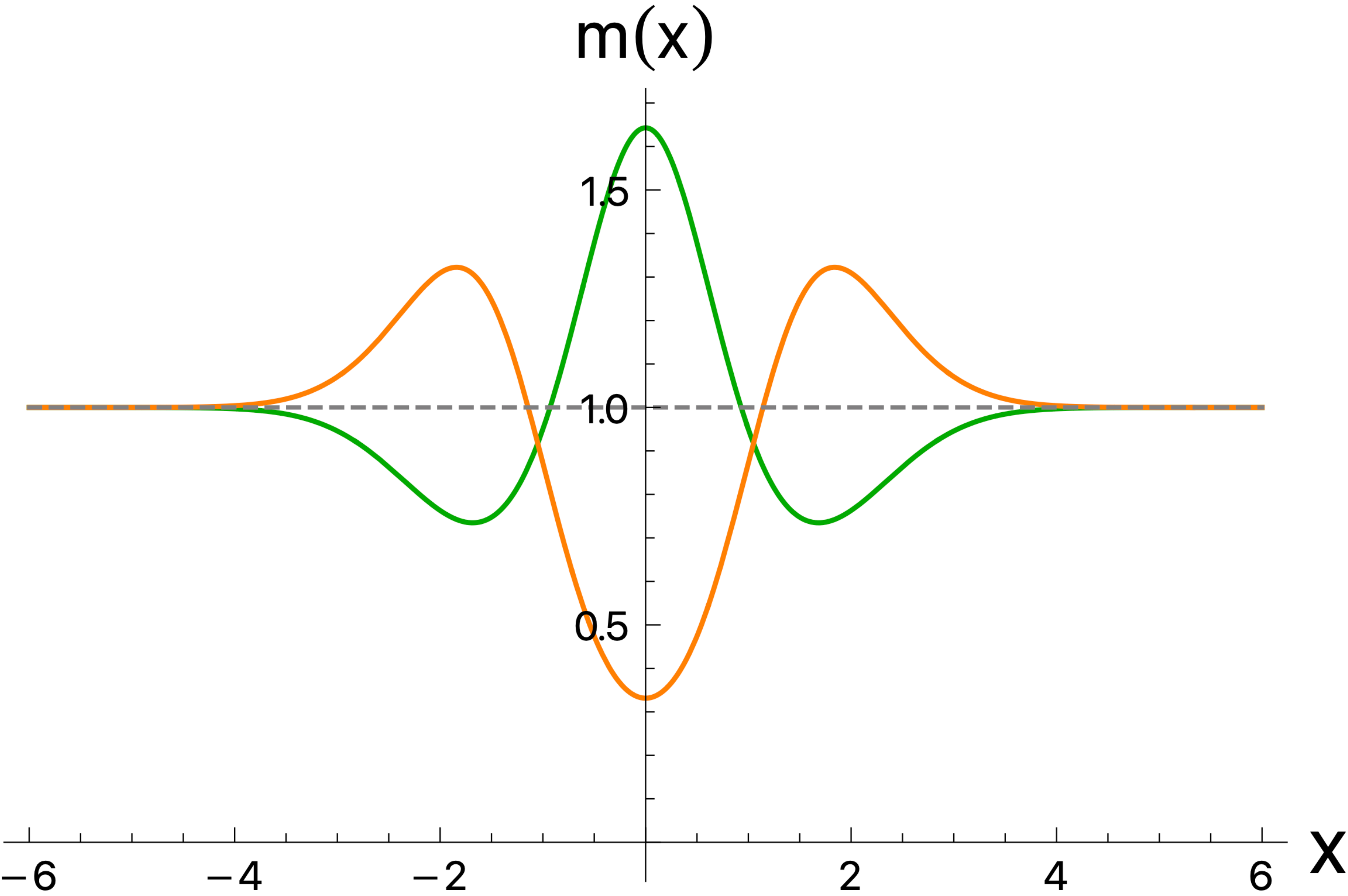}
    \includegraphics[width=7cm]{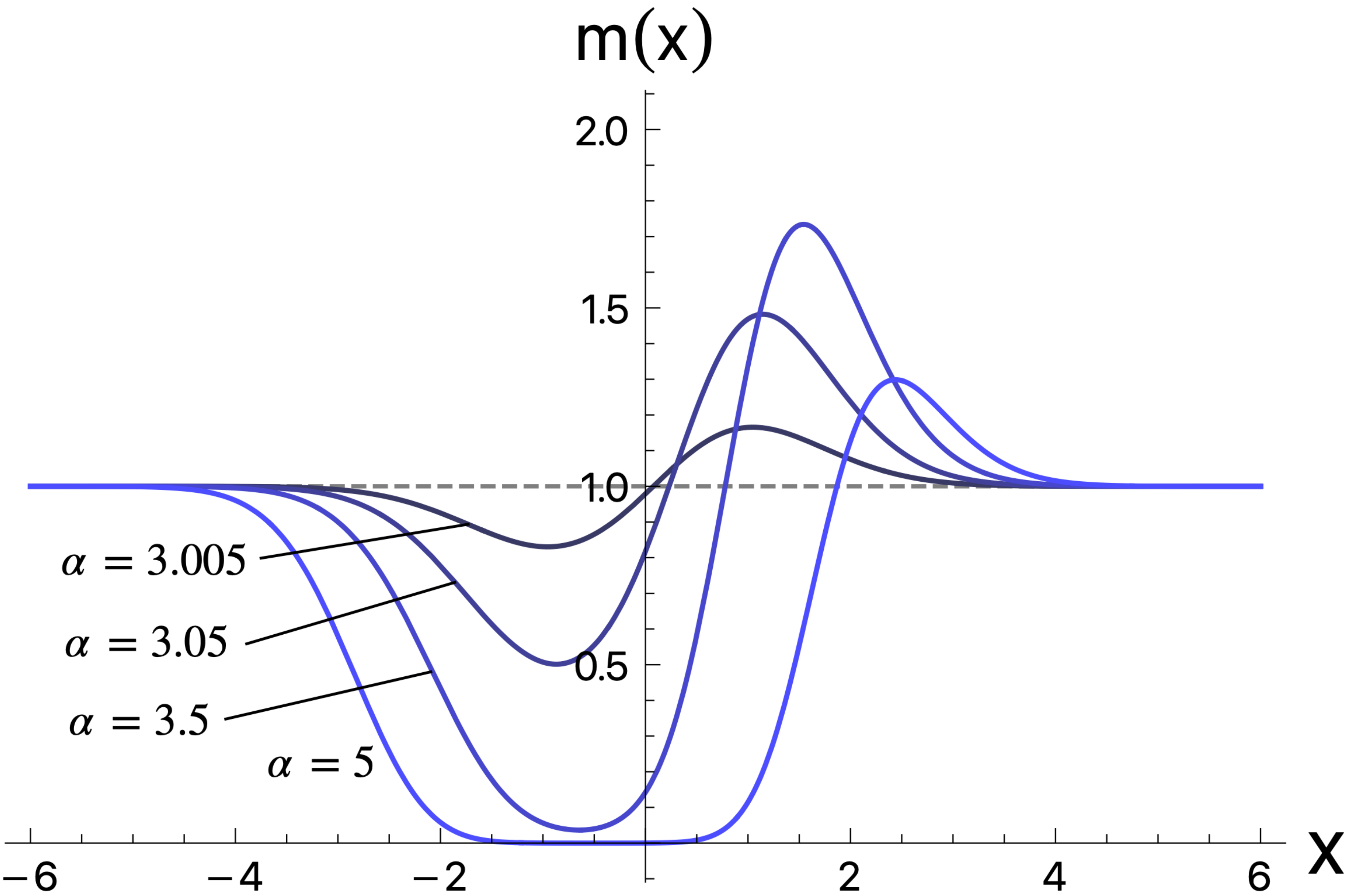}
    \caption{Solution profiles for deformed funnels. The above left panel shows the profiles for the first even branches, i.e., ditches ($m(0)<1$) and bulges ($m(0)>1$). The above right panel shows the second even branches at $\alpha=4.05$. The panel below is the first odd branch. Due to the small values of $m$, the profiles for large $\alpha=5$ resemble double droplets.}
    \label{fig:shape-funnels}
\end{figure}

To find funnel-type solutions, we resort to a Newton-Raphson algorithm with
\beq
    \cR(x) = (x^2+1)^\frac{\alpha-2}{2}e^{-x^2/2} f(x)\;,
\eeq
where we numerically solve for $f(x)$.  The phase diagram is shown in figure~\ref{fig:phasediagram-funnels} and \ref{fig:phasediagram-funnels-enlarge}.  Various solution profiles are shown in figure~~\ref{fig:shape-funnels}.  We see that the profile for large values of $\alpha$ resemble double droplets.


\paragraph{Stability of deformed funnels}
Figure~\ref{fig:phasediagram-funnels-enlarge} shows that the ditches have lower free energy than the uniform funnels, and hence more favorable than the uniform funnels, while the bulges have higher free energy and unfavorable. This thermodynamic instability of the bulges extends to the entire branch, as 
 an analytic approximation shows the bulge branch extends to $\Delta F \to \infty$ for $\alpha \to 1$  ( see Appendix~\ref{app:largebulges} ).
In figure~\ref{fig:Hessians-1steven}, we show various components of the Hesse matrices in eq.~(\ref{eq:def-Hij}), as well as its (partial) determinant.
These figures demonstrate the dynamical (linear) stability of the ditches with respect to several longer wavelength modes, as well as the instability of the bulges to the lowest even mode.

\begin{figure}
    \centering
    \includegraphics[width=8cm]{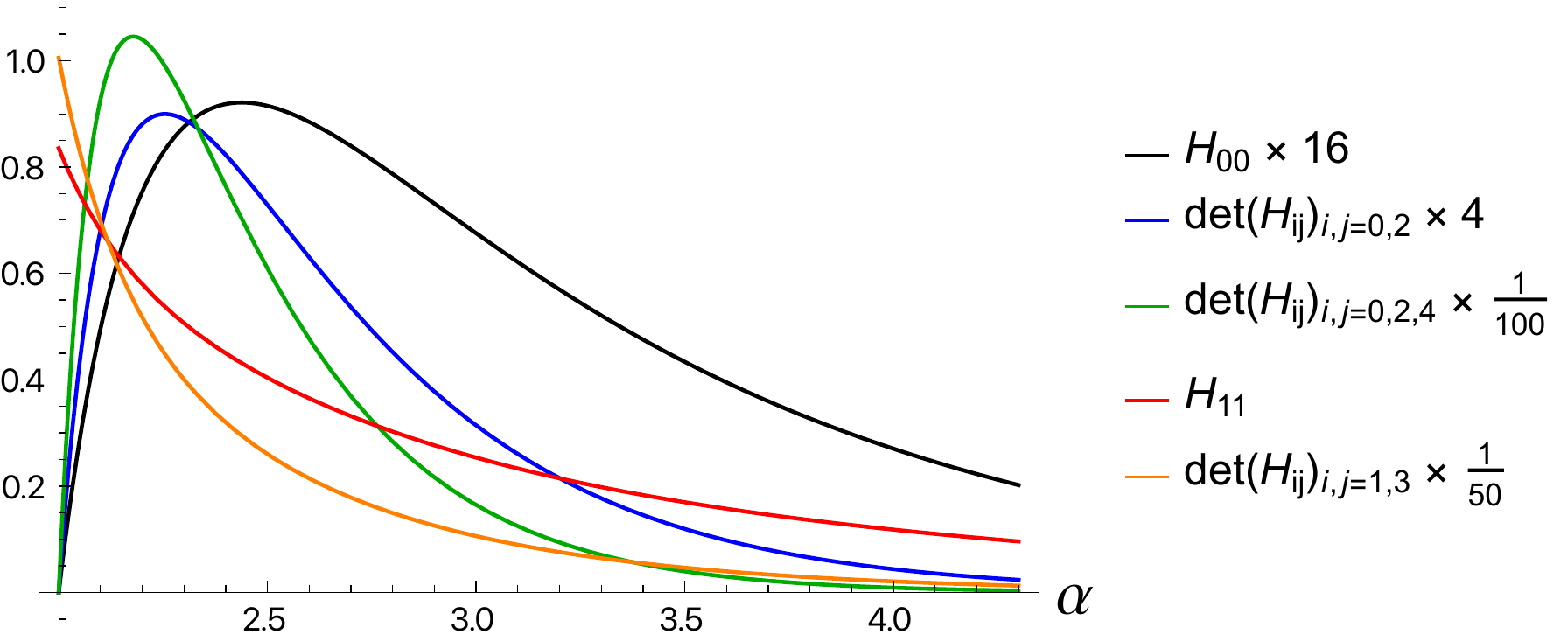}
    \includegraphics[width=7cm]{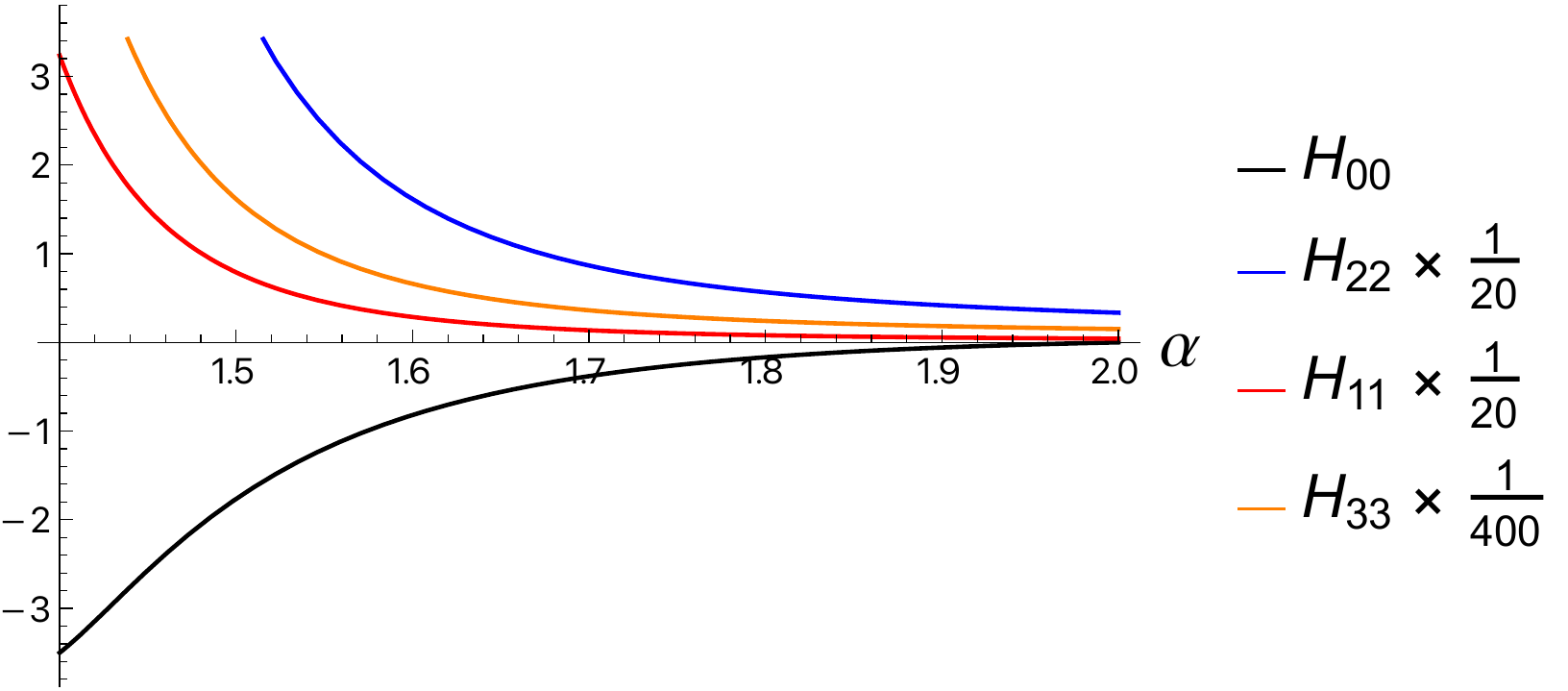}
    \caption{Hessians for ditches (left panel) and bulges (right panel). The left panel shows the linear stability of the ditches involving the first three even modes and first two odd modes. The right panel shows the bulges are unstable with respect to the first even modes.}
    \label{fig:Hessians-1steven}
\end{figure}

\subsection{Single Droplets}\label{sec:droplets}
\paragraph{Numerical scheme}  For single droplets, we use a shooting method.  In the left domain $(x<0)$, we suppose the solution is written in the following power series about $x=-\infty$ as
\beq
    \cR(x) =- \frac{\alpha+1}{2}x^2 +  \frac{\alpha+1}{\alpha-1} + C_L\, x^{1-\alpha^{-1}} \left(1+\ord{x^{-1},x^{-1-\alpha^{-1}}}\right),
    %\sum_{i,j=0}^\infty A^j c_{i,j} x^{-i} x^{-j(1+\alpha^{-1})}.
\eeq
where the constant $C_L$ determines the amplitude of the asymptotic behavior. With this expansion, we approximate the solution at $x=-x_1$ for some large value $x_1$. Numerically integrating from $x=-x_1$ to $x=0$ with the Runge-Kutta method, we obtain $\left.(\cR(0),\cR'(0))\right|_{\rm left}$ as a function of $C_L$.

On the other hand,  for the right domain $(x>0)$, we introduce the following variable compatible to the boundary behaivior~(\ref{eq:staticeq-bdry-bhv})
\beq
    \cR(x) = (x^2+1)^\frac{\alpha-2}{2}e^{-x^2/2} f_R(x)
\eeq
where $f_R(x)$ remains finite at $x\to\infty$, and then we write $C_R=f_R(\infty)$.
By using the Newton-Raphson method in the right domain for a given value of $C_R$,
the values at the origin $\left.(\cR(0),\cR'(0))\right|_{\rm right}=(f_R(0),f_R'(0))$ are obtained as a function of $C_R$.
Thus, the match at the origin determines $(C_L,C_R)$ for each $\alpha$
%( figure~\ref{fig:R0-dR0-plot})
\beq
    \left.(\cR(0),\cR'(0))\right|_{\rm left}(C_L) =     \left.(\cR(0),\cR'(0))\right|_{\rm right}(C_R).
\eeq

%\begin{figure}
%    \centering
%    \includegraphics[width=12cm]{plots/halfditchR0dR0_25_desc.pdf}
%    \caption{Plots of $(R(0),R'(0))$ solved in each domain with different parameters for $\alpha=2.5$. The droplet solutions are given by cross sections of the curves.
%    The cross sections between one of each and $R'(0)=0$ correspond to the Gaussian blob and even funnels.}
%    \label{fig:R0-dR0-plot}
%\end{figure}

\begin{figure}
    \centering
    \includegraphics[width=7cm]{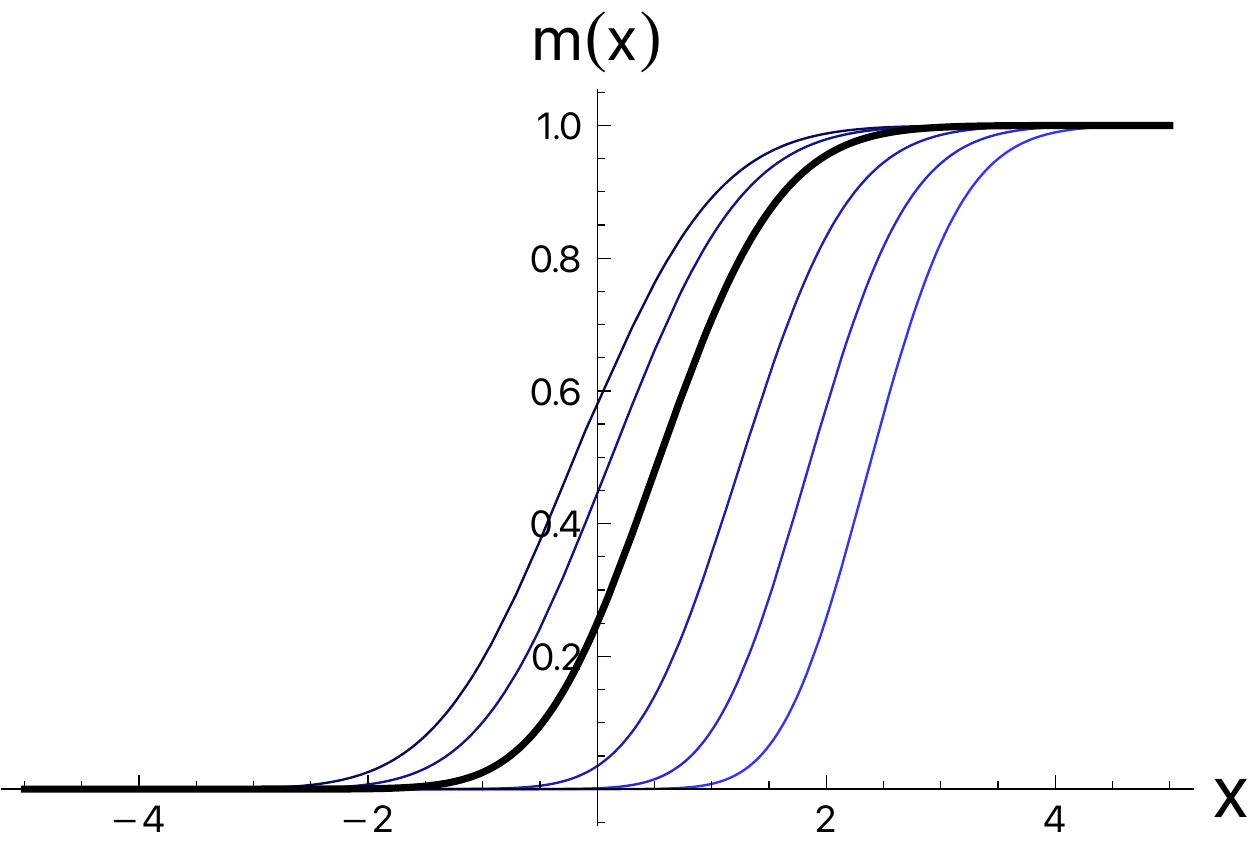}
    \,
    \includegraphics[width=7cm]{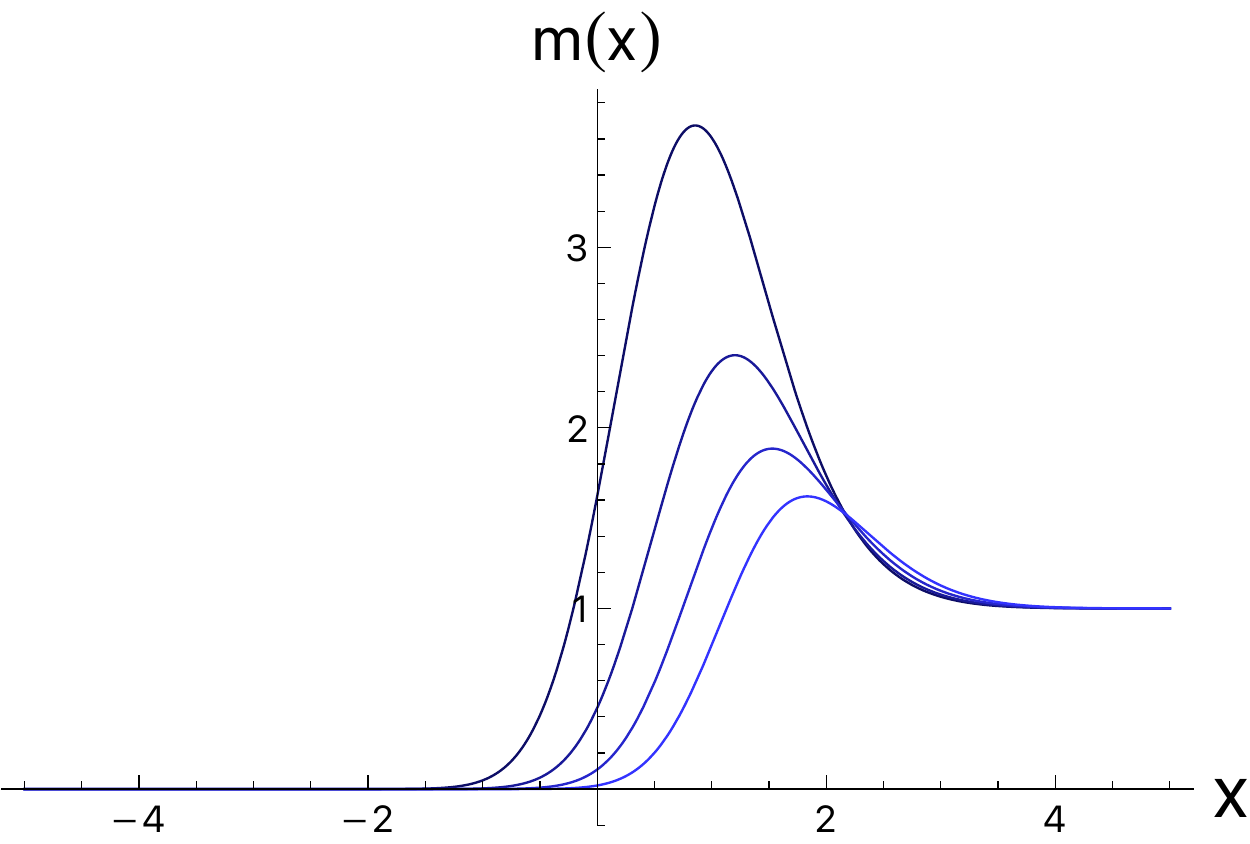}
    \caption{Solution profiles for droplets. The left panel shows the first droplet branch for $\alpha=0.2,0.5,1,2,3,4$ to the right. The thick black curve corresponds to the analytic solution with the error function for $\alpha=1$. The right panel shows the second droplet branch for $\alpha=2.5,3,3.5,4$ to the right. }
    \label{fig:shape-droplets}
\end{figure}

\paragraph{Analytic solution for $\alpha=1$}
For $\alpha=1$, there is an analytic solution to the the static equation~\eqref{eq:staticeq} given by
\beq
    \cR(x) = A+2\log\left[ B+{\rm err}\left(\frac{x}{\sqrt{2}}\right)\right].
\eeq
To satisfy the droplet boundary conditions, we must set $A=-2\log 2,\ B=1$ which leads to
\beq
    \cR(x) = 2 \log\left[\fr{2}\left(1+{\rm err}\left(\frac{x}{\sqrt{2}}\right)\right)\right],
\eeq
or
\beq
    m(x) = \fr{4}\left(1+{\rm err}\left(\frac{x}{\sqrt{2}}\right)\right)^2.
\eeq
In figure~\ref{fig:shape-droplets}, we present the profiles of the first and second droplets for several value of $\alpha$.

%--------------------------------------------------------------
\section{Horizon embeddings in global AdS: Soap bubble analysis}\label{sec:soapbubble}
%--------------------------------------------------------------

%The large $D$ effective theory that we have studied above, in fact, can be viewed as small fluctuations on the uniform funnels that deform the horizon radius in $\ord{1/D}$.
One can view the large $D$ dynamical effective theory as small fluctuations about a constant embedding of the horizon.  In our case, the horizon embedding is a constant  $r=r_0={\rm const}.$ surface in the vacuum AdS background
 \beq
    ds^2 = \frac{L^2}{\cos^2 z}\left(dz^2  - (\signK+r^2)d\tau^2+\frac{dr^2}{\signK+r^2}+r^2 d\Sigma_{n+1}^2\right),
 \eeq
where we also focus on the neck region of small $z\sim x/\sqrt n$.
Although the large $D$ effective theory can describe the nonlinear deformations of the horizon in terms of the mass function $m$, the change in the size remains $\ord{1/D}$ from $r=r_0$.

Instead, one can consider more general horizon embeddings $r= \bar{r}(z)$, which should satisfy the so-called soap bubble equation~\cite{Emparan:2015hwa}. For instance, horizon embeddings resembling AdS black droplets can be found by solving the soap bubble equation in the Poincar\'e AdS background~\cite{Emparan:2015hwa}. Unfortunately, many solutions to the soap bubble equation are only known numerically, which complicates the study of their dynamical fluctuations.  Nevertheless, solutions to the soap bubble equation provide useful information about the entire space of solutions.  Furthermore, some of the analytic solutions we have found can be interpreted within this soap-bubble analysis.

\subsection{Soap bubble equation.}
In what follows, we study the possible horizon shape in the $\signK=1$ case.  Consider the horizon embedding given by $r= \bar{r}(z)$ and the near-horizon coordinate $\rho$ given by
\beq
    r = \bar r(z)e^{\rho/n} = \bar{r}(z)(1+\rho/n+\ord{n^{-2}})
\eeq
Then, the trace of extrinsic curvature $K$ and the red-shift factor in the matching region $B$ $(0 \ll \rho \ll n)$ are given by
    \begin{align}
    K\bigr|_B \simeq  \pm \frac{n}{L} \frac{(1+ \bar{r}(z)^2)\cos z-  \bar{r}(z) \bar{r}'(z)\sin z}{ \bar{r}(z)\sqrt{1+ \bar{r}(z)^2+ \bar{r}'(z)^2}}
    \label{eq:meanK}
    \end{align}
and
    \begin{align}
        \sqrt{-g_{tt}}\bigr|_B \simeq \frac{L \sqrt{1+ \bar{r}(z)^2}}{\cos z}
    \end{align}
Note that for $r>\bar{r}(z)$ becomes the exterior of the horizon, the mean curvature $K\bigr|_B$ should be positive. In turn, The solution with $K\bigr|_B<0$ gives cavity-like shape~\cite{Emparan:2015hwa}.
These quantities should satisfy the soap bubble equation which is equivalent to the constraint equation
\beq
  \sqrt{-g_{tt}} K\bigr|_B =  2 \kappa.
\eeq
As the both sides are $\ord{n}$, we normalize the surface gravity as $k = \kappa/n$ and get
    \begin{align}
         \frac{\sqrt{1+ \bar{r}(z)^2}(1+ \bar{r}(z)^2- \bar{r}(z) \bar{r}'(z)\tan z)}{ \bar{r}(z)\sqrt{1+ \bar{r}(z)^2+ \bar{r}'(z)^2}} = 2k,\label{eq:soapbubble}
    \end{align}
where we chose $(+)$ signature in \eqref{eq:meanK} to include the uniform solution. The equation with $(-)$ signature is obtained by changing $k\to -k$, in which case $r<\bar{r}(z)$ is the exterior.

\subsection{Solutions.}
We obtain three basic types of solutions: uniform strings, droplets, and black holes.  By combining these, it is also possible to have multi-horizon solutions as long as the horizons do not intersect.  This is possible because separate horizons do not interact at infinite $D$.

\paragraph{Uniform funnels.}
Although most of the solutions of \eqref{eq:soapbubble} need numerical integration, several solutions admit analytic forms. The most trivial one is the uniform funnels $ \bar{r}(z)=r_0$, which satisfies
\beq
    r_0+\fr{r_0} = 2k
\eeq
or
\beq
    r_0 = k\pm \sqrt{k^2-1}.
\eeq
This shows that the surface gravity of uniform funnels have a lower bound $k \geq 1$. Note that the same condition holds for $ \bar{r}(z)$ if $ \bar{r}'(z)=0$ at any point. Conversely, $ \bar{r}'(z)$ never vanishes if $k \leq 1$, and hence $ \bar{r}(z)$ becomes a monotonic function.

\paragraph{AdS black holes.}
The soap bubble equation also admits AdS black holes as an analytic solution
    \begin{align}
         %\bar{r}(z) = \sqrt{A \cos^2 z-1},\quad A=2k(k\pm\sqrt{k^2-1}).
        \bar{r}(z) = \sqrt{(r_{H}^2+1) \cos^2 z-1},\quad k=\fr{2}\left(r_H+\fr{r_H}\right)
    \end{align}
    where the constant $r_H$ corresponds to the horizon radius.
    It is easy to check that this reproduces the Gaussian decaying behavior in the mass density for $z = x/\sqrt{n}$
    \begin{align}
        \frac{ \bar{r}(x/\sqrt{n})^n}{ \bar{r}(0)^n} \simeq e^{-\frac{1}{2}(1+r_H^{-2})x^2}\label{eq:adsbh-Gaussian-sb}
    \end{align}
    By identifying the radii of black holes and funnels $r_H=r_0$,  this behavior matches the asymptotic behavior of the Gaussian blob in the effective theory in the intermediate region $1\ll x \ll \sqrt{n}$.

\begin{figure}
\begin{center}
\includegraphics[width=11cm]{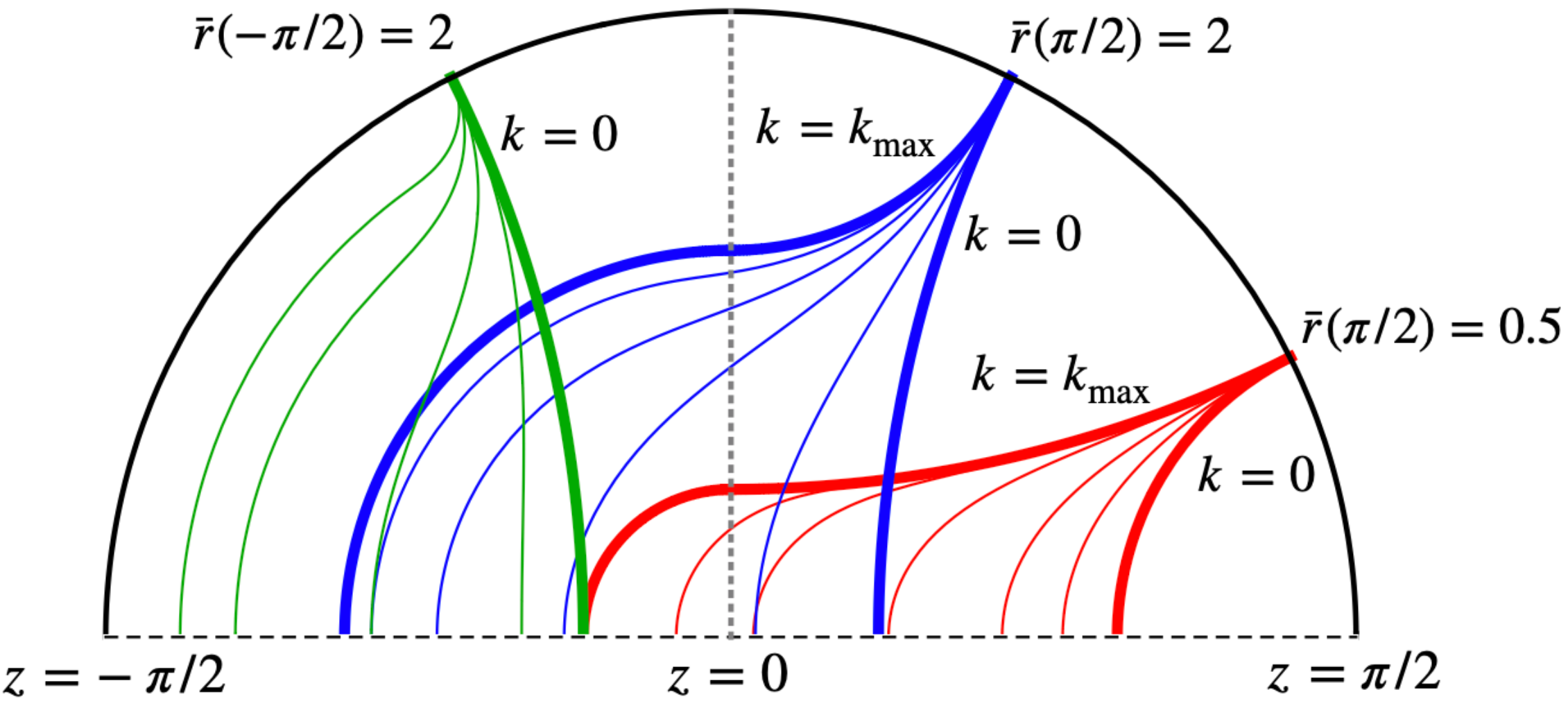}
\caption{Single droplets with different boundary radii embedded in the Poincar\'e disk. Each curve represents the horizon with its exterior to the left, and therefore, one can identify the blue and red curves as the single droplets in usual sense, while the green ones occupy more than the half of the disk, and hence looks more like a black cavity which is also found in the Poincar\'e AdS setup~\cite{Emparan:2015hwa}.  The blue and red branches have the minimum and maximum shapes which admit minimum and maximum value of the surface gravity respectively. In contrast, the green branch only has the minimum solution and admit arbitrary large surface gravity.  \label{fig:droplets_in_PD}}
\end{center}
\end{figure}

\paragraph{Droplets}
Except for a special case, the droplets can only be solved numerically as in figure~\ref{fig:droplets_in_PD}.  Here we set the boundary condition as $\bar{r}(\pi/2)=r_0$ and assume there exists a point $z = z_{\rm cap}$ such that $\bar{r}(z_{\rm cap})=0$. Depending on the signature of $K$, we have two type of branches for given $r_0$, both of which can be seen as droplets, but the negative branch rather looks like cavities. %For a given boundary value $r_0$, droplets with a maximum and minimum surface gravity have an analytic form.
Each branch has the minimum size at zero surface gravity $k=0$,
which admits an analytic form
\beq
    \bar{r}(z) = \sqrt{(r_0^2+1) \sin^2 z-1}.
\eeq
A similar zero surface gravity solution has also appeared in the Poincar\'e AdS case~\cite{Emparan:2015hwa}.

\begin{figure}
    \centering
    \includegraphics[width=14cm]{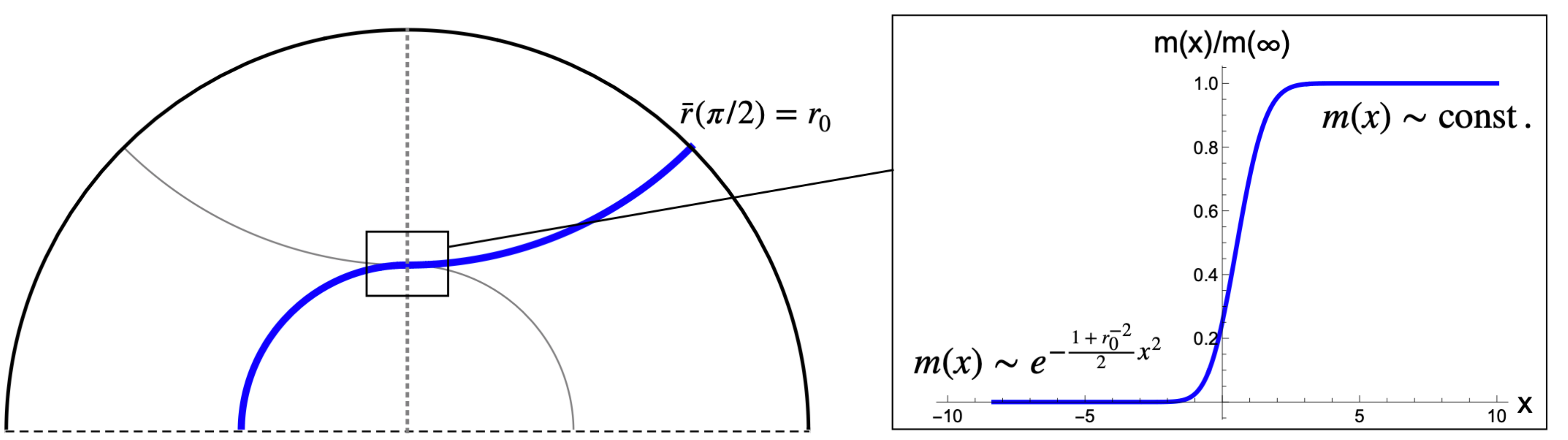}
    \caption{Relation between the maximum droplet in the soap bubble equation and single droplets in the effective theory.}
    \label{fig:glued_droplets}
\end{figure}

%\BW{Which values of $r_0$ admit maximum droplets, and which do not?}
The positive branches also has the maximum shape for given $r_0$.
The maximum droplet is made by gluing half of a uniform funnel and half of an AdS black hole at $z=0$, as shown in figure \ref{fig:glued_droplets}.
\beq
    \bar{r}(z) = \left\{ \begin{array}{cc}
    r_0     & (z>0) \\
    \sqrt{(r_0^2+1)\cos^2 z-1}     &  ( z<0 )
    \end{array} \right. .
\eeq
which has the maximum surface gravity of
\beq
    k_{\rm max} = \fr{2}\left(r_0 + \fr{r_0}\right).
\eeq
Although this gluing only admits $C^1$ continuity at $z=0$, this is still a solution of the soap bubble equation which is a first order ODE.  If we also zoom in on the region close to the origin by $z=x/\sqrt{n}$, we obtain a droplet-like solution to our effective theory described in section \ref{sec:droplets}.

The existence of the glued droplets implies a new aspect in the soap bubble analysis, which states that different soap bubble solutions can be cut and glued with $C^1$ continuity to make a new solution. %The $C^1$-joint will have a blow up in a large $D$ effective theory with corresponding asymptotics as we studied in the previous sections.

%It is unclear whether the $C^0$-joint is possible or not. At least, the fused conifold solution has the global geometry approximated by two joined black holes  in the large $D$ conifold analysis of Kaluza Klein black holes~[ref]. This suggests, if ever exists, the $C^0$-joint will not have a blow-up to the large $D$ effective theory, but rahter have a blow-up to large $D$ conifold geometry. In~[ESSTT15], the soap bubble equation in the Poincar\'e AdS background did not admit the black funnels. A possible scenario would be that the black funnels are obtained by allowing the $C^0$-joint between the uniform string and planer black brane.

%--------------------------------------------------------------
\section{Flowing funnels}\label{sec:flowingfunnels}
%--------------------------------------------------------------

As we have described, even when the boundary CFT spacetime contains multiple horizons, the bulk holographic dual may consist of a single connected horizon, and we have called these solutions ``black funnels."  But now suppose that we have two boundary black holes that are held at two different temperatures.  If the bulk description is still a black funnel, then it cannot have a globally well-defined temperature.  Similarly, if the two boundary black holes are allowed to rotate independently from each other, then the bulk horizon cannot rotate rigidly.  Such horizons are necessarily non-Killing, and are called "flowing horizons."  Flowing horizons do not violate any of the known black hole rigidity theorems by virtue of being non-compact.  Such horizons were first found in AdS \cite{Fischetti:2012ps,Fischetti:2012vt,Figueras:2012rb}, though see also \cite{Emparan:2013fha} for an example in flat space.

\paragraph{Steady flow setup}
We now construct flowing funnels using the effective large-$D$ equations. Until now, we have studied the static solution of eqs.~\eqref{eq:dtm} and \eqref{eq:dtmv}.
Here, instead, we consider the steady flow setup
\begin{align}
    \partial_t m = 0 ,\quad \partial_t v = 0,
\end{align}
in which, eq.~\eqref{eq:dtm} leads to
\beq
  \partial_x ( e^{x^2/2} mv) =  0 \quad \Longrightarrow \quad mv = P e^{-x^2/2},
\eeq
where $P$ is a constant which represents the mass flux through the boundaries.
Then, eq.~\eqref{eq:dtmv} results in the third order differential equation for $m(x)$,
\beq \label{eq:flowingf-m3}
  (\partial_x+x) \left(\frac{P^2 e^{-x^2}}{m} - 2m \partial_x \left(\frac{Pe^{-x^2/2}}{m}\right) - m \partial_x^2 \log m \right) -\alpha \partial_x m = 0.
\eeq

It is easy to see that the static equation~\eqref{eq:staticeq-m-3rd} is recovered by setting $P=0$.
This can be rewritten in terms of the local surface gravity~\eqref{eq:kappa-nlo} as
\beq
   \delta \kappa' = -\frac{Pe^{-x^2/2}(2m+xm'+m'')}{m^2}-\frac{2P^2e^{-x^2}(m'+xm)}{m},
\label{eq:flowingf-dkappa}
\eeq
where we introduced $\delta \kappa$ as the next to leading order terms in the local surface gravity~\eqref{eq:kappa-nlo}
\beq
\kappa =: 1 + \fr{n(\alpha+1)}\delta \kappa,
\eeq
which is written explicitly as
\beq \label{eq:flowingf-ddm}
  \delta \kappa = 2 - (\alpha-1)\log m - \frac{m''}{m}+\frac{m'^2}{2m^2}-\frac{xm'}{m}+
  \frac{P m' e^{-x^2/2}}{m^2} + \frac{3P^2 e^{-x^2}}{2m^2}.
\eeq
Eq.~\eqref{eq:flowingf-dkappa} shows that the surface gravity cannot be constant if $P\neq 0$, and hence, it has different boundary values, related to the mass function
\beq
\delta \kappa(-\infty) = 2 -(\alpha-1)\log m_L,\quad \delta \kappa(\infty) = 2 -(\alpha-1)\log m_R,
\eeq
where the boundary condition for $m(x)$ is given by
\beq
 m(-\infty) = m_L \neq m_R = m(\infty).
\eeq

To avoid the scaling degree of freedom under $m \to C m, P\to C P$, we define the relative mass flux compared with the left boundary mass
\begin{align}
    V := P/m_L.
\end{align}
The solutions are characterized by the difference in the boundary surface gravity, which is also scale invariant,
\beq
  \Delta \kappa := \delta \kappa(-\infty)-\delta \kappa(\infty) = (\alpha-1) \log (m_R/m_L),
\eeq
where $\Delta \kappa$ represents the fall of the surface gravity from the left to the right boundary.

By assuming $P \ll m(x)$, we obtain the linearized version of eq.~\eqref{eq:flowingf-m3} with $m(x)=m_0(1 + \delta m(x))$
\beq
    \delta m''' + x \delta m'' + \alpha \delta m' = 2 V e^{-x^2/2},
\eeq
where $V=P/m_0$. The solution is given by
\beq
    \delta m = \frac{\sqrt{2\pi}V}{\alpha-1} {\rm erf}\left(\frac{x}{\sqrt{2}}\right),\label{eq:flowingfunnel-linperturb}
\eeq
which estimates the relation between the flux and boundary surface gravity at the linear order,
\beq
    \Delta \kappa \simeq 2\sqrt{2\pi} V.\label{eq:flowingfunnel-lin-Vdk}
\eeq
This shows the energy flows from the hotter side to the cooler side.

At the nonlinear level, we used the Newton-Raphson method with the variables
\beq
   \cR(x):= \log (m(x)/m_L),\ {\cal Q}(x):= \cR'(x), \ {\cal K}(x) := \delta \kappa(x)-2+(\alpha-1)\log m_L
\eeq
which splits eq.~\eqref{eq:flowingf-m3} to the three first order equations
\begin{align}
&   {\cal K}' = V e^{-\frac{x^2}{2}-\cR} {\cal K}
+Ve^{-\frac{x^2}{2}-\cR} \left((\alpha -1)\cR-\frac{ {\cal Q}^2}{2}-2\right)-V^2 e^{-x^2-2\cR} \left(3 {\cal Q} +2 x \right)
   -\frac{3 V^3 e^{-\frac{3 x^2}{2}-3\cR}}{2},\\
 & {\cal Q}' =  2-{\cal K}-(\alpha -1) \cR+\frac{3}{2} V^2 e^{-2 \cR -x^2}
 - \left(x-V e^{-\cR -\frac{x^2}{2}}\right) {\cal Q}
 -\frac{{\cal Q} ^2}{2},\\
 & {\cal R}' = {\cal Q}.
\end{align}
The boundary condition is chosen so that
\beq
 \cR(-\infty) = 0,\quad \cR(\infty) = \Delta \kappa/(\alpha-1) ,\quad {\cal Q}(-\infty)=0={\cal Q}(\infty),\quad {\cal K}(-\infty) =0,\quad {\cal K} (\infty) = -\Delta \kappa.
\eeq

\begin{figure}
    \centering
    \includegraphics[width=7.2cm]{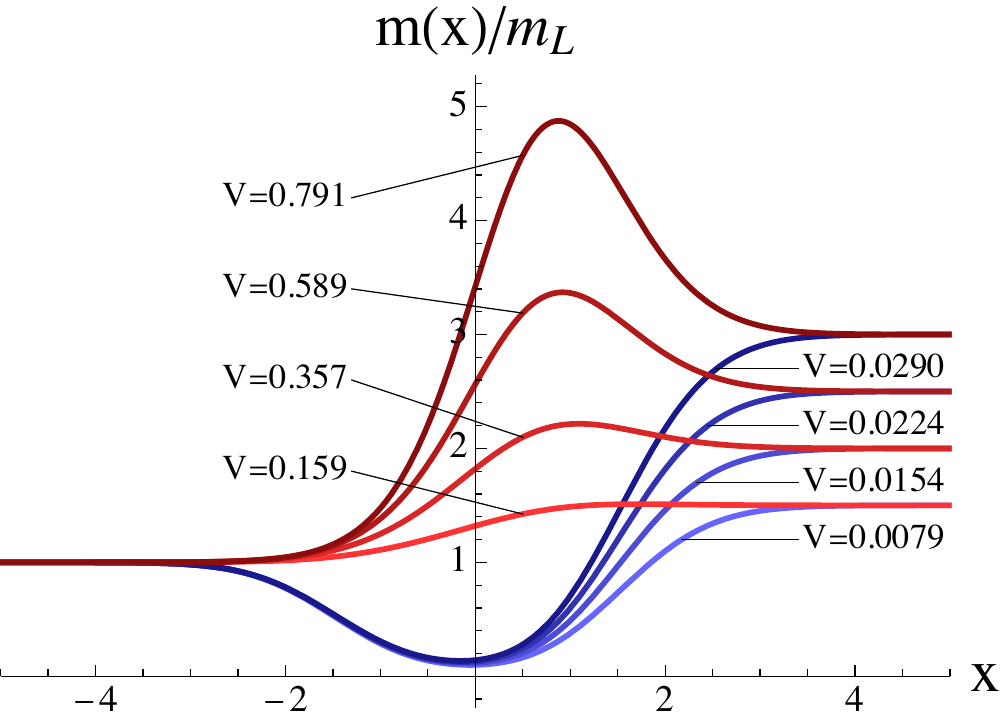}
    \hspace{0.1cm}
    \includegraphics[width=7.2cm]{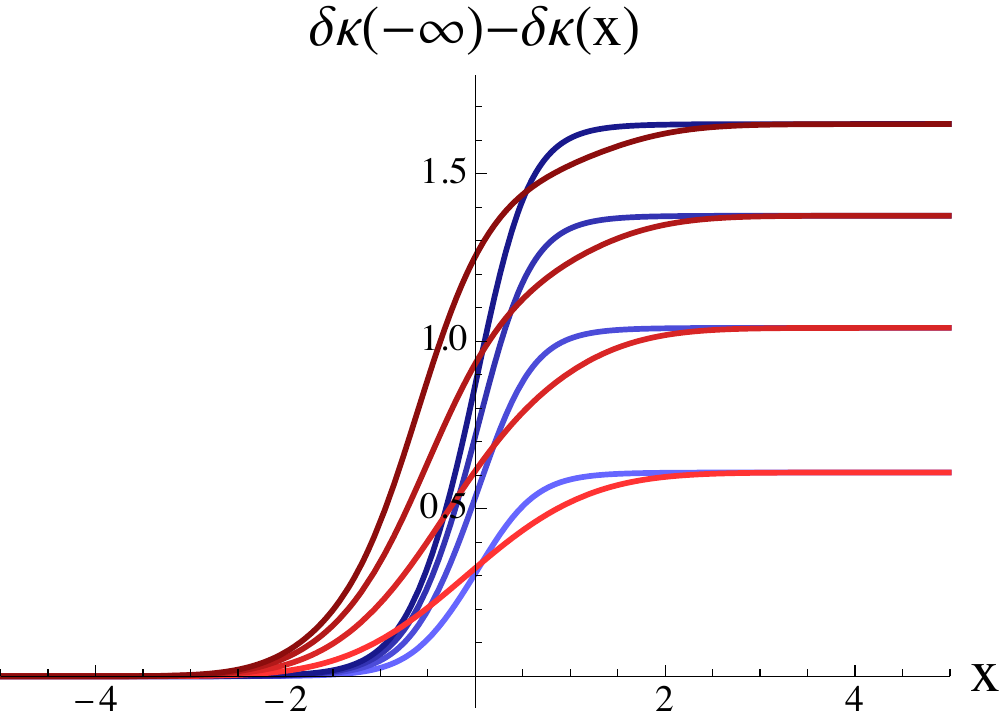}
    \caption{Flowing funnels for $\alpha=2.5$.
    Two distinct branches are plotted by red and blue curves.}
    \label{fig:flowingfunnel-shapes}
\end{figure}

\begin{figure}
    \centering
    \includegraphics[width=8cm]{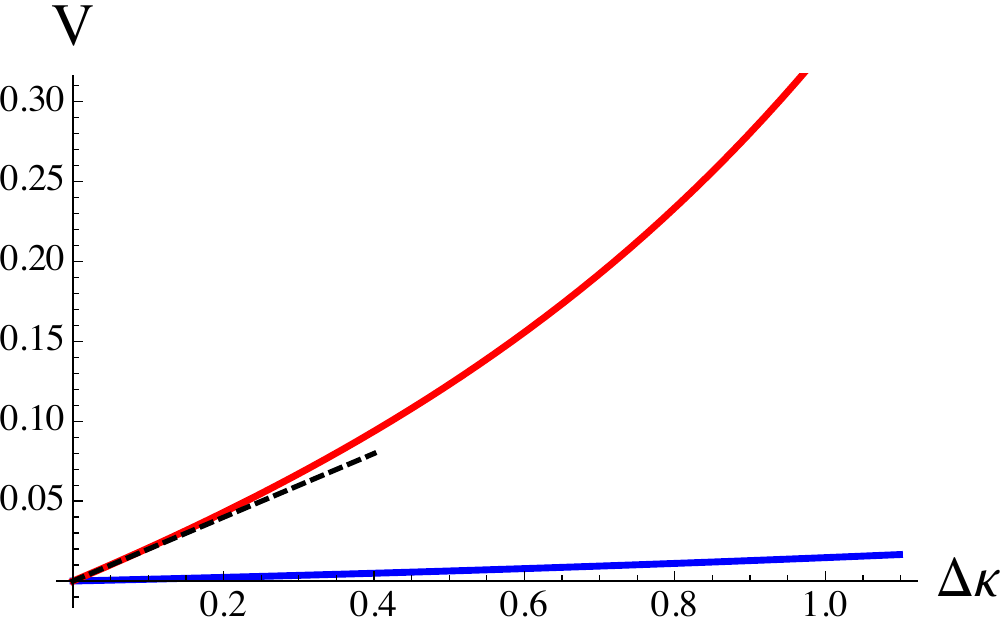}
    \caption{Relation between $\Delta \kappa$ and $V$ for $\alpha=2.5$ flowing funnels. Two thick curves correspond to two branches of the same colors in figure~\ref{fig:flowingfunnel-shapes}. The dashed line is given by the linear approximation~\eqref{eq:flowingfunnel-lin-Vdk}.}
    \label{fig:flowingfunnel-Vdk}
\end{figure}

In figure~\ref{fig:flowingfunnel-shapes}, we present the solutions for $\alpha=2.5$ with several different boundary conditions.
As in the static case, we have several deformed branches for the same $(\alpha,\Delta \kappa)$. In figure~\ref{fig:flowingfunnel-shapes}, one can find the red branches approaches to the uniform funnel at $V=0$, while the blue ones to the ditch. Interestingly, those branches admit different characteristics on the flux (figure~\ref{fig:flowingfunnel-Vdk}).
Due to the middle thin region which prevents the interaction between two boundaries, the blue branch admits much fewer flux than the red one.
On the other hand, the red branch admits large flux beyond the linear growth at the large temperature drop, which is caused by the bulge formation in the middle.
With the larger value of $\alpha$, one can expect more branches as seen in the static phases.


\subsection{Flowing funnels with shear flow}

By allowing an extra dimension to fluctuate, it is possible to construct a flowing horizon analytically.  Consider the black ``string" with $\signK =0$ and $r_0=1$, where the boundary CFT contains two planar black holes.  Boosting the black holes relative to each other will create a flowing horizon in the bulk.  This configuration has an analytic solution in a large $D$ effective theory.  Building off of the solution
\begin{equation}
ds^2=\frac{L^2}{\cos^2z}\left(dz^2-\left(1-\frac{1}{r^n}\right)dt^2+2dtdr+r^2dy^2+r^2d\Sigma_{n}\right)\;,
\end{equation}
we take the ansatz
\begin{equation}
ds^2=L^2\left[-Adt^2+2Udtdr+\frac{G_x}{n}\left(dx+G_cdy+\frac{C_x}{G_x}dt\right)^2+r^2\frac{G_y}{n}\left(dy+\frac{C_y}{G_y}dt\right)^2+r^2Sd\Sigma_n\right]\;,
\end{equation}
where the metric functions depend on $(t,x,y,\rho)$, where $x$ and $\rho$ are defined according to \eqref{eq:coordtransf}.  The metric functions have the large-$D$ expansion
\begin{align}
A&=1-e^{-\rho}m(t,x,y)\,,& C_x&=e^{-\rho}p_x(t,x,y)\,,& C_y&=e^{-\rho}p_y(t,x,y)\,,\\
G_x&=1+\left[\frac{x^2}{2}+\frac{e^{-\rho}p_x^2}{2m}\right]\frac{1}{n}\,,& G_y&=1+\left[x^2+\frac{e^{-\rho}p_y^2}{2m}\right]\frac{1}{n}\,& G_c&=\frac{e^{-\rho}p_xp_y}{2m}\frac{1}{n}\,,\\
U&=1+\left[x^2-\frac{e^{-\rho}(p_x^2+p_y^2)}{2m}\right]\frac{1}{n}\,,&S &= 1+\frac{x^2}{n}\;,
\end{align}
and satisfy the effective equations
\begin{subequations}
\begin{align}
    \partial_t m + (\partial_x +x) (p_x-\partial_x m)+\partial_y(p_y-\partial_y m)&=0\,,\\
    \partial_t p_x-(\partial_x +x)\left(\partial_x p_x-\frac{p_x^2}{m}\right)-\partial_y\left(\partial_yp_x-\frac{p_xp_y}{m}\right)+p_x-\partial_x m&=0\,,\\
    \partial_t p_y-(\partial_x +x)\left(\partial_x p_y-\frac{p_xp_y}{m}\right)-\partial_y\left(\partial_yp_y-\frac{p_y^2}{m}\right)+\partial_y m&=0\,.
\end{align}
A simple flowing solution is given by
\begin{equation}
m=m_0\;,\qquad p_x=0\;,\qquad p_y=V_0\,\mathrm{erf}\left(\tfrac{x}{\sqrt2}\right)\;.
\end{equation}
for some constants $m_0$ and $V_0$.
\end{subequations}

%--------------------------------------------------------------
\section{Summary}\label{sec:summary}
%--------------------------------------------------------------

In this article, we have studied various static phases as well as some steady flowing phases of black holes in the global AdS background using the large $D$ effective theory approach.

First, we have examined the dynamical and thermodynamic stability of static solutions deformed from the uniform string/funnel with both ends attached to the AdS boundaries.
These solutions are involved in the time evolution of the AdS black string instability that was studied in \cite{Emparan:2021ewh}.

We have derived the large $D$ effective theory of the deformed funnel by scaling up the central neck region by $\ord{\sqrt{D}}$ with different cross sections which contain the maximally symmetric space of the signature $K=1,0,-1$. 
We have found that the effective theory is parametrized uniquely by $\alpha:=K/r_0^2$ where $r_0$ is the funnel size. 

This effective theory admits the free energy functional which is monotonically decreasing in time.
Therefore, the solution with the lowest free energy must be stable at least linearly.
However, the conclusion is not so straightforward.
As discussed in \cite{Emparan:2021ewh}, the effective theory admits solutions with different types of boundary condition, the one is funnels which keep a finite radius at the AdS boundary, and the other is the Gaussian solution which has a Gaussian decay instead. The Gaussian solution can be interpreted as either of the AdS black black or very fat funnels which cannot be distinguished at large $D$. The difference in boundary behaviour produces an infinite difference in the free energy due to the AdS warped factor. Actually, it has shown that the Gaussian solution has infinitely lower free energy than any funnels for $\alpha>1$ and the opposite for $\alpha<1$. 
Such infinite difference cannot be seen from the perturbative study. 
Instead, it can lead to an indefinite growth in the time evolution as seen in \cite{Emparan:2021ewh}.

Then, we have analytically studied the thermodynamics and stability of the uniform funnels as well as the Gaussian blob by solving the 
effective equation perturbatively.
We have found that the Gaussian blob is linearly stable in any case.
While the uniform funnels admits the zero modes of the instabilities at $\alpha=k+2$, which implies the dynamical instability for $\alpha>2$.
For $\alpha<1$, since the Gaussian blob has the infinitely larger free energy than the funnels, the Gaussian blob is only metastable and the uniform funnel is absolutely stable. On the other hand for $\alpha>1$, the Gaussian blob is absolutely stable instead, and any linearly stable funnels are only metastable.

%The thermodynamics and stability of non-uniform solutions were studied in section~\ref{sec:pertnonuniformstab} for perturbative solutions and in section~\ref{sec:funnels} for numerical solutions.
We have also studied the phase and stability of non-uniform solutions branching from each zero mode $\alpha=k+2$ of the uniform solution both perturbatively and numerically.
In particular, from the lowest mode $k=0$ we have found two families of solutions which we have called {\it ditches} (where the center is caved) and {\it bulges} (where the center swells).  Our results indicate that bulges exist for $1<\alpha<2$, are unstable having higher free energy than the uniform solution.  On the other hand, ditches exist for $\alpha>2$, are metastable having lower free energy than the uniform solutions. 

Imposing the mixed boundary condition which assumes a finite radius on the one side and Gaussian decay on the other side, we have also obtained the single droplet solutions.
We have found that the droplet solutions have several deformed branches as the non-uniform funnels, which are solved numerically in general.

All of the above results are fully consistent with what was seen in a time evolution of this system in \cite{Emparan:2021ewh}.

By going to higher order in $1/D$, we have found the critical dimension where the non-uniform black strings switch in their direction in phase space \eqref{eq:static-criticalD} and in stability \eqref{eq:stabcriticalD}.  In other words, for $D\leq 11$, the bulges exist for $\alpha>\alpha_c$ (where $\alpha_c$ is the critical onset for the instability), are dynamically stable, and have lower free energy to the uniform solutions; while the ditches exist for $\alpha<\alpha_c$, are dynamically unstable, and have higher free energy to the uniform solutions.  For $D\geq 12$, the opposite occurs.  A similar kind of critical dimension is observed for asymptotically black strings \cite{Sorkin:2004qq,Suzuki:2015axa,Emparan:2018bmi}, where the critical dimension lies between $D=13$ and $D=14$.

Since, in our setup, the black funnel ends at two different positions in the AdS boundary, 
the boundary black holes do not necessarily have the same temperature.
Actually, we have found the large $D$ effective theory also admits some steady-state configurations, or {\it flowing solutions}, by imposing two slightly different boundary temperatures.
The flowing solutions have energy flux from the hotter side to the cooler side.
The construction of flowing solutions using the large $D$ effective equations was remarkably simple, especially given the difficulty of numerically constructing these solutions at finite $D$.  This provides an ideal setting to study the steady-state transport properties of horizons, especially out of equilibrium. 
Indeed, the large $D$ effective theory has been applied to a non-equilibrium steady-state configuration
to study the time evolution of the discontinuity in the initial state, i.e. the Riemann problem in the planer AdS brane setup~\cite{Herzog:2016hob}. 
We leave this study for future work.

Finally, we have explored more general horizon embeddings in the global AdS beyond the regime of the effective theory formulated around the central neck region. The embedding condition at large $D$ is given by the so-called soap bubble equation, which we have solved numerically in most cases.
As the result, we have obtained different types of droplets which cannot be described by the previous effective theory.
Moreover, we have also found that certain droplets can be obtained by gluing parts of a funnel and a black hole with a $C^1$-joint, which corresponds to a single droplet solution in the effective theory. This type of solution has not been found in the soap bubble analysis in the asymptotically flat or Poincar\'e AdS background~\cite{Emparan:2015gva}. One might ask about solutions with a $C^0$-joint. Although such solutions fail to satisfy the soap bubble equation at the joint, they can be considered weak solutions %if the corresponding solution in the effective theory is smooth.
if they have a smooth continuation around the joint. Such continuations are not necessarily described by the large $D$ effective theory.
For example, if two droplets are tangent, the near-joint shape approaches to the double cone geometry~\cite{Kol:2002xz,Kol:2003ja,Asnin:2006ip,Emparan:2011ve}. At large $D$, in particular, it is shown that the cone-like (conifold) ansatz describes the topology-changing transition which actually asymptotes to the two tangent
black holes~\cite{Emparan:2019obu}\footnote{The large $D$ conifold analysis is also applied to the Einstein-Gauss-Bonnet theory~\cite{Nair:2021lxz}.}.

Dynamical effective equations will be obtained by considering small (in $1/D$) fluctuations about the soap-bubble solutions, but this is often not feasible when the equation is only known numerically.  To expand the range of applicability of large $D$ methods, it would be useful to develop a more general large $D$ formalism where such effective equations can be obtained, even if the solutions to the soap bubble equation are only known numerically.

Throughout the article, we have considered the black funnels/droplets which ends at the AdS boundary.
Alternatively, one can place a brane at finite distance in the bulk, where a boundary condition can be imposed instead.
This setup might be useful for understanding, for instance, black hole evaporation in braneworld models~\cite{Emparan:2022toappear}. 

\section*{Acknowledgement}
We thank Roberto Emparan for useful comments and discussions.
This work is supported by ERC Advanced Grand GravBHs-692851.
DL is supported by a Minerva Fellowship of the Minerva Stiftung Gesellschaft fuer die Forschung mbH.
RS is supported by JSPS KAKENHI Grant Number JP18K13541 and
partly supported by Osaka City University Advanced Mathematical Institute (MEXT Joint Usage/Research Center on Mathematics and Theoretical Physics)
as well as Toyota Technological Institute Fund for Research Promotion A.
BW acknowledges support from MEC grant FPA2016-76005-C2-2-P.

%--Appendix---------------------------------------------------
\appendix

%--------------------------------------------------------------
\section{Large \texorpdfstring{$D$}{D} effective theory for long wavelengths}\label{app:longwavelength}
%--------------------------------------------------------------

In the main part, we consider the large $D$ effective theory of short wavelength $\Or{1/\sqrt{n}}$, which is crucial to
capture the instability and its non-linear growth properly. This scaling is actually consistent with the perturbative result~\cite{Marolf:2019wkz} (see also Appendix~\ref{app:marolfsantos}).
Here we show that, without rescaling, the large $D$ limit leads to an alternative effective theory which does not admit the instability.
We start with the following ansatz
\beq\label{eq:metans2}
ds^2=\frac{L^2}{\cos^2z}\left[ H dz^2 -A dt^2+2 u_t dt dr-\frac{2}{n} C dt dz +r^2 d\Omega_{n+1}\right]\;,
\eeq
which differs from \eqref{eq:metans} only in that the coordinate $z$ has not been scaled with $D$.  This is a different large $D$ limit and should capture different physics of the black string. For simplicity, we only consider $K=1$ case.

Setting $r=r_0e^{\rho/n}$ and taking the large $n$ limit, we find at lowest order
\begin{align}\label{eq:ACHu2}
    A&=(1+r_0^2)[1-e^{-\rho}m(t,z)]\,,& C&=r_0e^{-\rho}p(t,z)\,,\\
    H&=1+e^{-\rho}\frac{r_0^2p^2(t,x)}{n^2(1+r_0^{2}) m(t,x)}\,,& u_t&=1\,,
\end{align}
where $m$ and $p$ satisfy the dynamical equations
\begin{equation}
    \partial_t m + \tan z(p-\partial_z m)=0\,,
\end{equation}
\begin{equation}
    \partial_t p-\tan z\left(\partial_zp-\frac{p^2}{m}\right)+(1-\tan  z^2)p-\left(1+\alpha\right)\partial_z m=0\,.\label{eq:longwavelength-momeq}
\end{equation}
where we introduced $\alpha:=r_0^{-2}$ and the time is rescaled by $t \to t/r_0$ as in the main part.

%The black string is given by the solution $m=1$, $p=0$.
%Seeking time independent perturbations about this solution, we find
%\begin{equation}
%    m=1+\epsilon \left[(\sin z)^{1-\frac{1}{r_0^2}}-1\right]\;,\qquad p=\epsilon \left(1+\frac{1}{r_0^2}\right)\cos z(\sin z)^{-1/r_0^2}\;,
%\end{equation}
%which satisfies the effective equations to $O(\epsilon)$, and the condition that $m(t,\pm\pi/2)=1$.  We see that there is no solution that is regular at $z=0$, so there is no zero mode to an instability.
To find the static configuration, we need $p=\partial_z m$ and eq.~\eqref{eq:longwavelength-momeq} leads to
    \beq
    \tan z m''-\tan z \frac{(m')^2}{m} + (\alpha+\tan^2z)m'= 0.
    \eeq
This is solved analytically to give
    \beq
        m(z) = m_0 \exp( C \sin^{1-\alpha} z)
    \eeq
    and
    \beq
        p(z) = m_0 C (1-\alpha) \cos z \sin^{-\alpha} z \exp(C\sin^{1-\alpha} z).
    \eeq
We see that there is no static solution regular at $z=0$ expect the uniform string $C=0$.
This shows the uniform string has no zero mode to an instability. 

It is not clear whether solutions with $C\neq 0$ are unphysical or not. 
    For $\alpha>1$, the exponent is divergent at $z=0$ and might have connections to funnels or droplets type solutions, depending on the signature of $C$.
    While for $0<\alpha<1$, we always have non-divergent fat/thin strings which have a kink at $z=0$. This kink can be blown up to the blobology solution with the alternative boundary condition in eq.~\eqref{eq:staticeq-bdry-bhv}
    \beq
    \sin^{1-\alpha} z \simeq n^\frac{1-\alpha}{2} (x/\sqrt{\alpha})^{1-\alpha}\quad {\rm for} \quad 1 \ll x := \sqrt{n} z \ll \sqrt{n}.
    \eeq

%--------------------------------------------------------------
\section{Different coordinates for AdS and its isometries} \label{app:embeddingAndBoostingOfAdS}
%--------------------------------------------------------------

Throughout this paper, we will use different coordinates to describe the background geometry in $AdS_D$, which are suitable for each horizon configuration.
An easy way to understand the relation between those coordinates is to consider the embedding of AdS$_D$ as a hyperboloid in the $D+1$ two-timing flat space
\beq
ds^2=-dX_0^2-dX_{-1}^2+dX_{-2}^2+dX_1^2+\dots dX_{n+2}^2\,,
\label{eq:MinkDplus2}
\eeq
with the embedding condition
\beq
-L^2=-X_0^2-X_{-1}^2+X_{-2}^2+X_1^2+\dots X_{n+2}^2\,,
\label{eq:AdSHyperb}
\eeq
where $n:=D-4$.
In the following, we present several background coordinates used in this paper.
\paragraph{Global AdS}
\beq
ds^2 = L^2 \left( -(1+R^2)dT^2+\frac{dR^2}{1+R^2}+R^2 (d \theta^2+\sin^2\theta d\Omega_{n+1}^2)\right),
\eeq
which is obtained by the embedding
\beq
 X_{-2} = L R\cos\theta ,\quad X_{-1}=L \sqrt{1+R^2}\cos T,\quad X_0=L \sqrt{1+R^2} \sin T \quad X_{i\geq 1} = L R \sin\theta \omega_i
\eeq
where
\beq
\sum_{i=1}^{n+2}\omega_i^2=1.
\eeq
This is convenient to describe the spherical AdS black hole
\beq\label{eq:schads}
ds^2=L^2\lp -F(R)dT^2+\frac{dR^2}{F(R)}+R^2\lp d\theta^2+\sin^2\theta d\Omega_{n+1} \rp\rp\,,
\eeq
where
\beq
F(R)=1+R^2-\frac{\mu}{R^{n+1}}\,.
\eeq

\paragraph{Poincar\'e disk}
When we draw the static horizon embedded in the global AdS, it is handy if we embedded it in the spacial conformaly flat disk region, so called Poincar\'e disk,
\beq
  ds^2\bigr|_{T={\rm const}} = \frac{4L^2}{(1-\tilde{R}^2)^2}\left(
  d\tilde{R}^2 + \tilde{R}^2 (d\theta^2+\cos^2\theta d\Omega^2_{n+1})\right),
\eeq
where $0\leq \tilde{R}<1$.
The coordinate $\tilde{R}$ is related to the radial coordinate in the global AdS by
\beq
   R = \frac{2\tilde{R}}{1-\tilde{R}^2}.
\eeq

\paragraph{AdS$_{D-1}$-slicing with $S^{n+1}$}

\beq\label{eq:adsz}
ds^2 =\frac{L^2}{\cos^2 z}\lp dz^2 +ds_{D-1}^2\rp
\eeq
where $-\pi/2 <z <\pi/2$ and
\beq
  ds_{D-1}^2= -\lp 1+r^2\rp dT^2+\lp 1+r^2\rp^{-1} dr^2+r^2 d\Omega_{n+1},
\eeq
with
\beq
X_0=\frac{L \sqrt{1+r^2} \cos T}{\cos z},\quad
 X_{-1}= \frac{L \sqrt{1+r^2} \sin T}{\cos z},\quad
X_{-2}= L\tan z,\quad X_{i\geq 1}=\frac{Lr}{\cos z}\,\omega_i,
\eeq
This coordinate is convenient for the AdS black string
\beq\label{eq:adsstring}
ds^2_{D-1}=-f(r) dT^2+f(r)^{-1} dr^2+r^2 d\Omega_{n+1}
\eeq
where
\beq
f(r)=1+r^2-\frac{\mu}{r^n}\,.
\eeq
The coordinates~(\ref{eq:adsz}) are mapped to the Poincar\'e disk by
\beq \label{eq:adsbs-to-poincare}
 (\tilde{R} \cos\theta,\tilde{R} \sin\theta) =\left( \frac{\sin z}{\sqrt{1+r^2} + \cos z}, \frac{r}{\sqrt{1+r^2}+\cos z}\right).
\eeq

\subsection{Symmetries}
Now, we show that a symmetric transformation in the hyperboloid invokes a boost transformation in the effective theory. Let us consider the following transformation
\begin{align}
&   \tilde{X}_0 = \cosh \beta  X_0 - \sinh \beta X_{-2} \\
&   \tilde{X}_{-2} = -\sinh \beta X_0 + \cosh \beta X_{-2}\\
&   \tilde{X}_{-1} = X_{-1},\quad \tilde{X}_{i\geq 1} = X_{i\geq 1}
\end{align}
which is rewritten in the embedded coordinate~(\ref{eq:adsz}) as
\begin{align}
    \frac{\cos \tilde{T} \sqrt{1+\tilde{r}^2} }{\cos \tilde{z}} &=\frac{\cosh \beta  \cos T \sqrt{1+r^2}}{\cos z}-\sinh \beta  \tan z\,, \label{eq:trafo1}\\
    \tan \tilde{z} &=\cosh \beta  \tan z-\frac{\sinh \beta  \cos T \sqrt{1+r^2}}{\cos z}\,, \label{eq:trafo2}\\
    \frac{\sin \tilde{T}\sqrt{1+\tilde{r}^2} }{\cos \tilde{z}} &=\frac{\sin t \sqrt{1+r^2}}{\cos z}\,.  \label{eq:trafo3}\\
    \frac{\tilde{\omega}_i \tilde{r} }{\cos \tilde{z}}&=\frac{\omega_i  r}{\cos z}\,.
    \label{eq:trafo4}
\end{align}
Using eqs.~(\ref{eq:trafo1})-(\ref{eq:trafo3}), it is easy to show
\beq
    \frac{\tilde{r}}{\cos \tilde{z}} = \frac{r}{\cos z},
\eeq
and hence $\tilde{\omega_i}=\omega_i$ as well.
Since the transformation of other coordinates $(T,r,z)$ in general is complicated, we instead consider the transformation in the $1/D$ expansion with the large $D$ coordinate near $r=r_0$ surface, \footnote{For simplicity, we do not rescale the time $t\to t/r_0$ as in the main part.}
\begin{align}
    t&= T + \arctan(r_0\, e^{\rho/n})-\arctan(r_0)\,,\\
    x &= \sqrt{n} z \,,\\
%    \sR &=  (r/r_0)^n\,,
    e^\rho &=  (r/r_0)^n\,,
\end{align}
where $t$ denotes the time in the Eddington-Finkelstein coordinate. The large $D$ coordinates after the transformation $(\tilde{t},\tilde{\rho},\tilde{x})$ is introduced by the same formula with $(\tilde{T},\tilde{r},\tilde{z})$.
We also assume the boost parameter is of ${\cal O}(1/\sqrt{n})$ by rescaling as
\beq
    \beta \to \frac{\beta}{\sqrt{n}\sqrt{1+r_0^2}},
\eeq
where the scaling factor is introduced for later convenience.
Then, the transformation in the large $D$ coordinate is obtained by expanding in $1/n$,
\begin{align}
&\tilde{t} = t - \frac{\beta(2x-\beta \cos t)(r_0 \cos t-\sin t)}{2(1+r_0^2)n}+\Or{n^{-2}},\\
&\tilde{x} = x-\beta \cos t+\Or{n^{-1}},\\
%&\tilde{\sR} = \sR \left(\frac{\cos \tilde{z}}{\cos z}\right)^n =
%\sR \exp\left(\fr{2} (x^2-\tilde{x}^2)\right)+\Or{n^{-1}}.\nonumber\\
%&\qquad = \sR \exp\left(x \beta \cos t-\fr{2}\beta^2 \cos^2t\right)+\Or{n^{-1}}.
&e^{\tilde{\rho}} = e^\rho \left(\frac{\cos \tilde{z}}{\cos z}\right)^n =
e^\rho \exp\left(\fr{2} (x^2-\tilde{x}^2)\right)+\Or{n^{-1}}.\nonumber\\
&\qquad = e^\rho \exp\left(x \beta \cos t-\fr{2}\beta^2 \cos^2t\right)+\Or{n^{-1}}.
\end{align}
Since the leading order metric~(\ref{eq:metans}) with \eqref{eq:ACHu}
share the same background geometry before and after the transformation,
given the solution in $(t,x)$, the following becomes the solution in $(\tilde{t},\tilde{x})$
\begin{align}
    m(t,x) &\longrightarrow \tilde{m}(\tilde{t},\tilde{x})
    = m(\tilde{t},\tilde{x}+\beta \cos \tilde{t}) \exp \left(\tilde{x} \beta  \cos \tilde{t} + \fr{2} \beta^2 \cos^2 \tilde{t}\right)\,, \\
    v(t,x) &\longrightarrow \tilde{v}(\tilde{t},\tilde{x})=v(\tilde{t},\tilde{x}+\beta \cos \tilde{t})+\frac{\beta}{r_0} \sin \tilde{t}\,,
\end{align}
where the velocity field $v$ is defined by $p= mv + \partial_x m$. Taking into account the time rescaling $t\to t/r_0$, this gives the symmetry transformation in the effective theory in eq.~(\ref{eq:boostsym}) with $\signK=1$.

%--------------------------------------------------------------
\section{Large \texorpdfstring{$D$}{D} limit of the static deformation on the uniform funnel}\label{app:marolfsantos}
%--------------------------------------------------------------
Here we discuss some implication from the perturbative result in~\cite{Marolf:2019wkz}.
Following \cite{Marolf:2019wkz}, the $z$-dependence of the static perturbation regular at $z=\pm \pi/2$ is given by
\footnote{A different onset in $z$ where the boundary is at $z=0,\pi$ was used in~\cite{Marolf:2019wkz}.}
\beq
Y_\pm(z) = \hgfunc{2}{1}\left(-\beta,n+5+\beta+\frac{n}{2}+3;\frac{1\mp\sin z}{2} \right)\cos^{n+2}z,
\label{eq:static-mode-MS}
\eeq
where we set $D=n+4$ and
\beq
\beta:=-\frac{n+5}{2} + \sqrt{\frac{(n+3)^2}{4}+\Delta} \simeq \Delta/n -1+{\cal O}(n^{-1}),
\eeq
where $\Delta$ is a sepration constant.
Here $\beta$ should take a non-negative integer to impose the regularity at both sides, which imposes
\begin{align}
    \hat{\Delta}:=\Delta/n = k + 1+{\cal O}(n^{-1}),\quad k =0,1,2,\dots.
    \label{eq:MS-reg1}
\end{align}
Note that this is the exact solution without the large $D$ limit. Due to the warped factor $\cos^{n+2}z$, there is no proper large $D$ limit of $Y_\pm(z)$ over the entire region $-\pi/2 \leq z \leq \pi/2$.
The factor $\cos^{n+2}z$ remains finite at large $D$ only if we focus on the small central region by introducing $z=x/\sqrt{n}$ as we do in the effective theory analysis in section~\ref{sec:largeDeffectivetheory}. In other regions away from the neck, the large $D$ limit gives non-perturbatively small deformation which cannot be treated by the effective theory approach.

The radial dependence, in contrast, can be solved by using the usual $1/D$ expansion as performed in~\cite{Marolf:2019wkz}. The regularity on the horizon requires $\hat{\Delta}:=\Delta/n\simeq \alpha-1$ where $\alpha$ is given by eq.~\eqref{eq:def-alpha}. Along with the regularity condition for $Y_\pm(z)$
in eq.~\eqref{eq:MS-reg1}, we can conclude that the onset of the static deformation exists only if $\alpha \simeq k + 2+\Or{n^{-1}}$ for a non-negative integer $k$.

If $\beta$ does not take a non-negative integer, $Y_\pm(z)$ are no longer regular at the opposite side. This, however, does not mean the regular deformation is absent. Instead, we can suppose the perturbations from both sides pick up some information from our blobology result by using the matched asymptotic expansion near the central neck.
Using the following identities,
\begin{align}
&\hgfunc{2}{1}(a,b,c;x) = (1-x)^{-a} \hgfunc{2}{1}\left(a,c-b,c;\frac{x}{x-1}\right),\quad (x<1),\\
&\hgfunc{2}{1}(a,b,a-b+1;x) = (1+x)^{-a} \hgfunc{2}{1}\left(\frac{a}{2},\frac{a+1}{2},a-b+1;\frac{4x}{(1+x)^2}\right),\quad (|x|<1),
\end{align}
the mode function~(\ref{eq:static-mode-MS}) is transformed to
\beq
Y_\pm(z) =(\pm\sin z)^{\beta} \hgfunc{2}{1}\left(-\frac{\beta}{2},-\frac{\beta+1}{2},\frac{n}{2}+3;\pm\cot^2 z\right)\cos^{n+2} z.
\eeq
From the asymptotic behavior of $\hgfunc{2}{1}(a,b,c+\lambda;z)$ at $\lambda \to \infty$, we obtain
\beq
Y_\pm(z) \simeq (\pm \sin z)^{\beta} \left(1+{\cal O}\left(\cot^2 z/n\right)\right) \cos^{n+2}z.
\eeq
The correction terms can be ignored if $\cot^2 z \ll n$.
In the coordinate $z=x/\sqrt{n}$, this corresponds to the matching condition
\beq
  1 \ll |x| \ll \sqrt{n},
\eeq
and hence, the following match is available
\beq
Y_\pm(z) \simeq n^{1-\frac{\alpha}{2}}|x|^{\alpha-2}e^{-\frac{x^2}{2}},
\eeq
where the radial condition $\beta \simeq \hat{\Delta}-1 \simeq \alpha-2$ is used.
This is proportional to the asymptotic behavior of the static configuration at $x\to \pm \infty$ in the large $D$ effective theory~\eqref{eq:staticeq-bdry-bhv}.

%--------------------------------------------------------------
\section{Blob and neck construction of large bulges}\label{app:largebulges}
\begin{figure}[ht]
\begin{center}
\includegraphics[width=11cm]{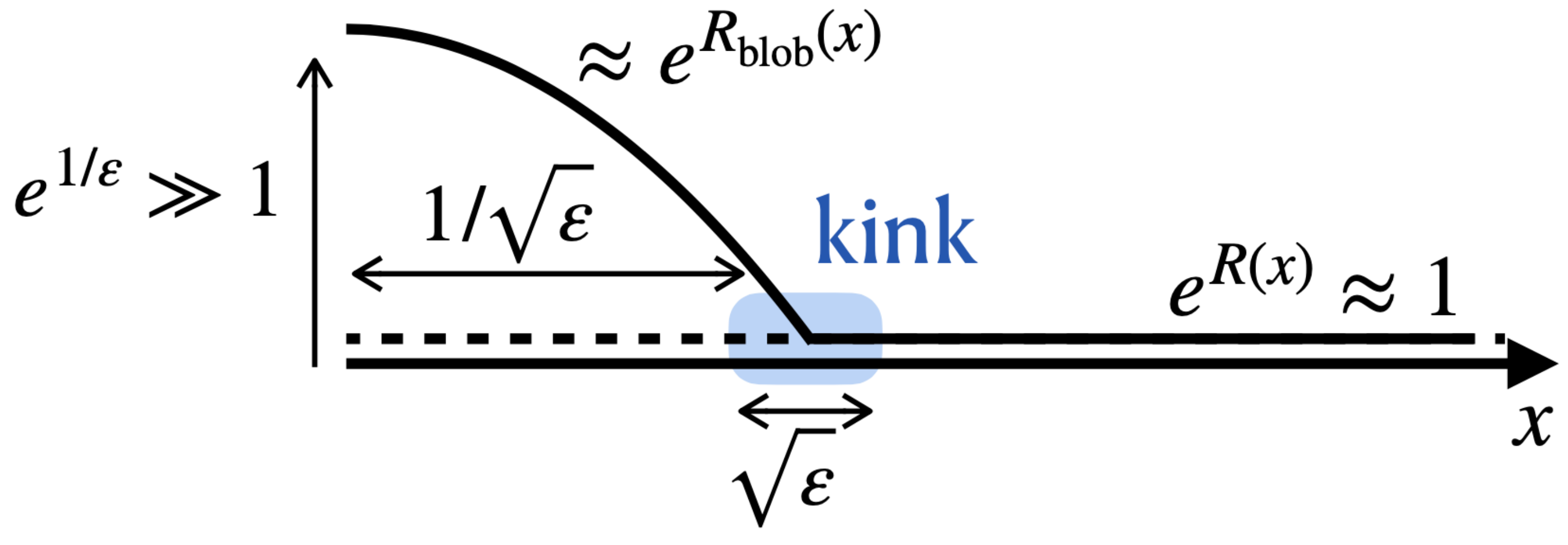}
\caption{Blob and kink in the bulge solution for $\alpha = 1+2\varepsilon$.}
\label{fig:bulge-blob-kink}
\end{center}
\end{figure}

%--------------------------------------------------------------
Here we present the analytic construction of the bulge solution close to $\alpha=1$ by using the blob and neck formalism~\cite{Suzuki:2020kpx}.
We solve the static equation with the fixed scale $\cR_0=0$
\begin{equation}
 \cR''(x)+\frac{1}{2}\cR'(x)^2+(\alpha-1)\cR(x)+x\cR'(x)=0.
\end{equation}
One can read off the solution shape for $\alpha \approx 1$ from the numerical solutions of bulges in section~\ref{sec:numericalphase},
which indicate that in the central region the solution is approximated by the large Gaussian blob
\begin{equation}
 \cR_{\rm blob}(x) = \frac{\alpha+1}{\alpha-1} - \frac{\alpha+1}{2} x^2,
 \label{eq:gaussianblob-blobkink}
\end{equation}
while in the asymptotic region the solution reduces to the uniform funnel $\cR(x)=0$. The transition between the two regions takes place in a very short wavelength compare to the width of the central blob, which produces a  sudden kink in the profile~(figure \ref{fig:bulge-blob-kink}). Hence, we expect a version of blob and neck construction~\cite{Suzuki:2020kpx}, or {\it blob and kink} construction is applicable with the assumption
\beq
 \alpha=1+2\varepsilon,\quad \varepsilon \ll 1.
\eeq
 In the following, we consider only the right half domain $x\geq 0$ as the bulge branch has the even parity.

The kink occurs where the blob and uniform solutions intersect, i.e. $x \sim \sqrt{2}(\alpha-1)^{-1/2}$. To resolve the intersecting region, we introduce a small scale coordinate $z$ near the kink
\begin{equation}
x = \frac{1}{\sqrt{\varepsilon}}+\sqrt{\varepsilon} \, z
 \label{eq:near-kink-coord}
\end{equation}
at which the Gaussian blob~(\ref{eq:gaussianblob-blobkink}) is expanded as
\begin{equation}
\cR_{\rm blob}(x)  \simeq -2z-\varepsilon (z^2+2z)+{\cal O}(\varepsilon^2).\label{eq:blob-kink-blobmatch}
\end{equation}

\paragraph{Near-kink solution}
In the near-kink coordinate~(\ref{eq:near-kink-coord}), the static equation becomes
\begin{equation}
 \cR''(z)+\frac{1}{2}\cR'(z)^2+\cR'(z)=- \varepsilon z \cR'(z) + {\cal O}(\varepsilon^2).
\end{equation}
The solution is obtained in the series expansion of $\varepsilon$,
\begin{equation}
\cR_{\rm kink}(z) = - 2 z + 2 \log (e^z+c_1) - \varepsilon \frac{c_1 z(z+2)}{e^z+c_1} + {\cal O}(\varepsilon^2),
\end{equation}
where $c_1$ is an undetermined constant to be matched and $\cR_{\rm kink}(\infty) = 0$ is also imposed so that the solution matches with the uniform funnel at $z\to \infty$. The expansion for $z\to \infty$ gives the matching condition with the uniform solution
\begin{equation}
 \cR_{\rm kink}(z) = e^{-z}(2-\varepsilon(2z+z^2)) + \ord{\varepsilon^2,e^{-2z}}.\label{eq:blob-kink-kinkmatch-as}
\end{equation}
For $z\to -\infty$, we obtain the matching condition with the central blob
\begin{align}
\cR_{\rm kink}(z) = - 2z + 2\log c_1  - \varepsilon (2 z +z^2)+ \frac{2e^z}{c_1} \left(1+\varepsilon \left(\frac{z^2}{2}+z\right)  \right) +{\cal O}(\varepsilon^2,e^{2z}).\label{eq:blob-kink-kinkmatch}
\end{align}
Comparing with the Gaussian blob~(\ref{eq:blob-kink-blobmatch}), the integration constant is determined as
\begin{equation}
c_1=1+{\cal O}(\varepsilon^2).
\end{equation}
As seen below, the existence of the small kink structure distorts the two connecting regions by small deformations.

\paragraph{Back reaction to Gaussian blob}
First, we consider the perturbation to the Gaussian blob
\begin{equation}
\cR(x)=\cR_{\rm blob}(x)+\delta \cR_{\rm blob}(x),
\end{equation}
which is given by
\begin{equation}
\delta \cR_{\rm blob}(x) = A_0 \,  H_{1-\fr{\alpha}}\left(\sqrt{\frac{\alpha}{2}}x\right)+ B_0\,  H_{1-\fr{\alpha}}\left(-\sqrt{\frac{\alpha}{2}}x\right)\label{eq:blob-kink-dblob-sol}
\end{equation}
where the even parity sets $A_0=B_0$.
%To determine $A_0$, we have to match with the kink solution.
%For the match, we use the asymptotic expansion at $x \to \infty$,
%\begin{align}
%& H_{1-\fr{\alpha}}\left(\sqrt{\frac{\alpha}{2}}x\right) = \ord{x^{1-\fr{\alpha}}},\nonum
%&H_{1-\fr{\alpha}}\left(-\sqrt{\frac{\alpha}{2}}x\right) =
%-\frac{\sqrt{\pi}(1-1/\alpha)}{\Gamma(1/\alpha)}\left( \sqrt{\frac{\alpha}{2}} x\right)^{-2+\fr{\alpha}} e^{\frac{\alpha x^2}{2} }\left(1+\frac{(1-2\alpha)(1-3\alpha)}{2\alpha^3 x^2}+\ord{x^{-4}}\right).
%\end{align}
The amplitude is determined through the match in the kink region~(\ref{eq:near-kink-coord}), in which eq.~(\ref{eq:blob-kink-dblob-sol}) is expanded as
\begin{align}
\delta \cR_{\rm blob}(x)
\approx -2\sqrt{2\pi} \,e\, \varepsilon^{3/2}\, e^\frac{1}{2\varepsilon} A_0\, e^{z}\left(1+\varepsilon \left(\frac{z^2}{2}+z-2-2\gamma+\log (2\varepsilon) \right)+{\cal O}(\varepsilon^2)\right),
\end{align}
where $\gamma$ is Euler's constant.
Comparison with ${\cal O}(e^z)$ terms in eq.~(\ref{eq:blob-kink-kinkmatch}) sets
\begin{equation}
A_0 = -(2\pi)^{-\fr{2}} \varepsilon^{-\frac{3}{2}} e^{-1-\frac{1}{2\varepsilon}}\left(1+\varepsilon \left(2+2\gamma-\log(2\varepsilon)\right)\right).
\end{equation}

\paragraph{Back reaction to asymptotic behavior}
Similarly, the asymptotic behavior at $x\to \infty$ is matched in terms of a small perturbation from the uniform solution $\cR(x)=0$
\begin{equation}
 \delta \cR_{\rm uni}(x)  = A_\infty \, e^{-\frac{x^2}{2}} H_{\alpha-2} \left(\frac{x}{\sqrt{2}}\right)+B_\infty \, e^{-\frac{x^2}{2}} H_{\alpha-2} \left(-\frac{x}{\sqrt{2}}\right).
\end{equation}
In the kink region~(\ref{eq:near-kink-coord}), this perturbation can be expanded as
\begin{equation}
\delta \cR_{\rm uni}(x) \simeq A_\infty \sqrt{\frac{\varepsilon}{2}}e^{-\frac{1}{2\varepsilon}}e^{-z}\left(1-\varepsilon \left(1+z+\frac{z^2}{2}+\log (\varepsilon/2)\right) \right)
+B_\infty \sqrt{\pi} \left(1+\varepsilon(2\gamma-\log (2 \varepsilon)\right).
\end{equation}
The comparison with eq.~(\ref{eq:blob-kink-kinkmatch-as}) determines
\begin{equation}
A_\infty = 2 \,e^\frac{1}{2\varepsilon} \sqrt{\frac{2}{\varepsilon}}\left(1+\varepsilon(1+\log(\varepsilon/2)\right),\quad B_\infty=0.
\end{equation}

\begin{figure}
    \centering
    \includegraphics[width=8cm]{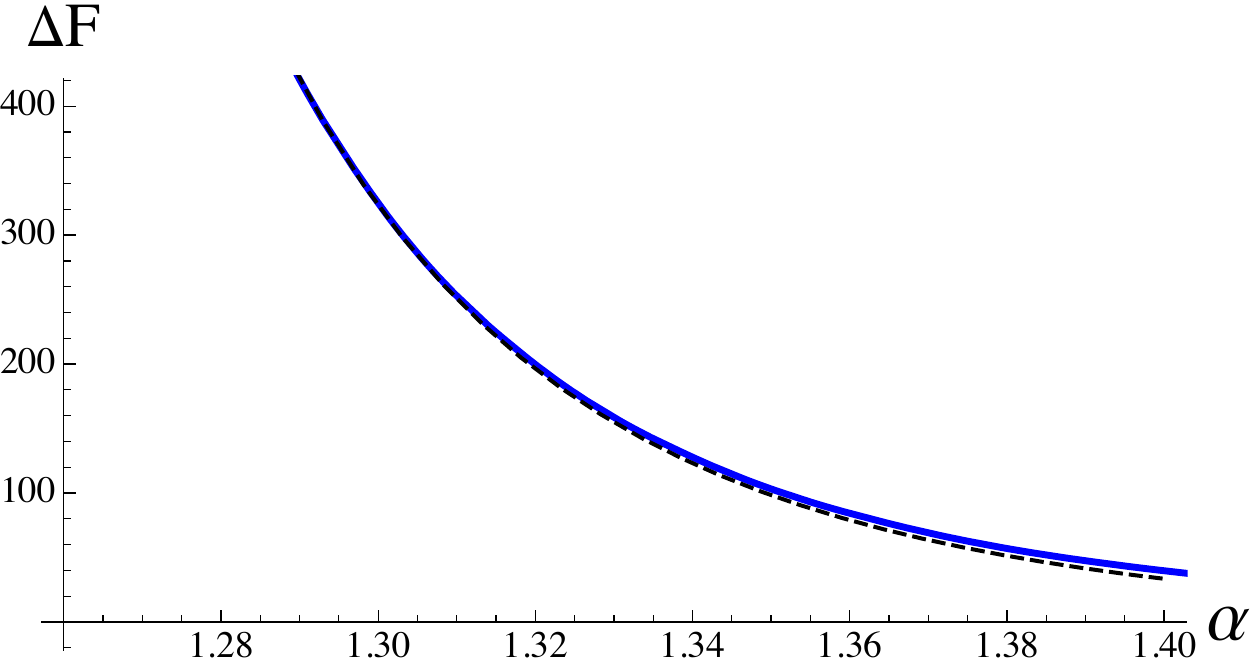}
    \caption{Bulge branches at large deformation. The dashed curve is the approximation by the blob and kink construction~(\ref{eq:largebulge-dF-formula})}
    \label{fig:phase-bulgeL}
\end{figure}

\paragraph{Free energy}
Combining the solutions in all regions, we can evaluate the free energy by dividing the integral into corresponding parts
\beq
    \int_0^\infty \Delta \rho_F dx= \int_0^{\fr{\sqrt{\varepsilon}}-\sqrt{\varepsilon} z_1}\Delta \rho_F\bigr|_{\rm blob}dx
+    \int_{\fr{\sqrt{\varepsilon}}-\sqrt{\varepsilon} z_1}^{\fr{\sqrt{\varepsilon}}+\sqrt{\varepsilon} z_2} \Delta \rho_F\bigr|_{\rm kink}dx
+    \int_{\fr{\sqrt{\varepsilon}}+\sqrt{\varepsilon} z_2}^\infty \Delta \rho_F\bigr|_{\rm asym}dx
\eeq
where $\Delta \rho_F = \rho_F - \rho_F\bigr|_{\rm uniform}$ and the cut-offs $z_1$ and $z_2$ should not appear in the final result. Thus, rewriting in terms of $\alpha$, we obtain
\beq
 \Delta F = \frac{\alpha-1}{\alpha+1}\sqrt{\frac{2\pi}{\alpha}}e^\frac{\alpha+1}{\alpha-1} - \frac{4\sqrt{2}}{\sqrt{\alpha-1}}e^\fr{\alpha-1}\left(1+\ord{(\alpha-1)^2}\right),\label{eq:largebulge-dF-formula}
\eeq
where the first term of $\ord{e^{1/\varepsilon}}$ is identical to the free energy (without subtraction) of the Gaussian blob which takes no further corrections in $\ord{\varepsilon}$, while the second term of $\ord{e^{1/2\varepsilon}}$ is the expansion up to $\ord{\varepsilon}$.
In figure~\ref{fig:phase-bulgeL}, this formula reproduces the phase diagram of bulges sufficiently close to $\alpha=1$.

\bibliography{refs}{}
\bibliographystyle{JHEP}

\end{document}